\documentclass[%
reprint,
superscriptaddress,
nofootinbib,
nobibnotes,
 amsmath,amssymb,
 aps,
 prd,
floatfix,
]{revtex4-2}

\usepackage{amsmath}
\usepackage[caption=false]{subfig}
\usepackage{booktabs}
\usepackage{rotating}
\usepackage{multirow} 
\usepackage{graphicx}% Include figure files
\usepackage{dcolumn}% Align table columns on decimal point
\usepackage{bm}% bold math
\usepackage[table,dvipsnames]{xcolor}
\definecolor{myblue}{RGB}{62,73,173}
\definecolor{myred}{RGB}{216, 28, 56} 
\usepackage[colorlinks=true,breaklinks=true,hyperfootnotes=true,allcolors=myblue]{hyperref}
\usepackage[nameinlink]{cleveref}
\usepackage{todonotes}
\usepackage{orcidlink}

\Crefname{section}{Sec.}{Sec.}
\Crefname{figure}{Fig.}{Figs.}
\Crefname{table}{Tab.}{Tabs.}

\usepackage{physics}
\usepackage{siunitx}
\AtBeginDocument{\RenewCommandCopy\qty\SI}

\renewcommand{\qc}{\, \text{,}}
\newcommand{\qs}{\, \text{.}}
\graphicspath{{fig/}}
\usepackage[english]{babel}

\usepackage{acro}
\acsetup{
barriers/use = true,
barriers/reset = true,
}

\DeclareAcroEnding{possessive}{'s}{'s}
\NewAcroCommand\acg{m}{\acropossessive\UseAcroTemplate{first}{#1}}
\NewAcroCommand\acsg{m}{\acropossessive\UseAcroTemplate{short}{#1}}
\NewAcroCommand\aclg{m}{\acropossessive\UseAcroTemplate{long}{#1}}
\NewAcroCommand\acfg{m}{%
    \acrofull
    \acropossessive
    \UseAcroTemplate{first}{#1}%
}
\NewAcroCommand\iacsg{m}{%
    \acroindefinite
    \acropossessive
    \UseAcroTemplate{short}{#1}%
}
\DeclareAcronym{AI}{
    short = AI,
    long  = artificial intelligence
}

\DeclareAcronym{AK}{
    short = AK,
    long  = analytic kludge
}
\DeclareAcronym{AAK}{
    short = AAK,
    long  = argumented analytic kludge
}
\DeclareAcronym{AUC}{
    short = AUC,
    long  = area under the curve
}
\DeclareAcronym{BH}{
    short = BH,
    long  = black hole
}
\DeclareAcronym{BHB}{
    short = BHB,
    long  = black hole binary,
    long-plural-form = black hole binaries
}
\DeclareAcronym{BBH}{
    short = BBH,
    long  = binary black hole
}

\DeclareAcronym{BNS}{
    short = BNS,
    long  = binary neutron star
}
\DeclareAcronym{CNN}{
    short = CNN,
    long  = convolutional neural network
}
\DeclareAcronym{CE}{
    short = CE,
    long  = Cosmic Explorer
}
\DeclareAcronym{CBC}{
    short = CBC,
    long  = compact binary coalescence
}
\DeclareAcronym{CBS}{
    short = CBS,
    long  = compact binary system
}

\DeclareAcronym{DECODE}{
    short = DECODE,
    long  = DECODE
}
\DeclareAcronym{DECIGO}{
    short = DECIGO,
    long  = DECi-hertz Interferometer Gravitational wave Observatory
}

\DeclareAcronym{DNN}{
    short = DNN,
    long  = deep neural network
}
\DeclareAcronym{ESA}{
    short = ESA,
    long  = European Space Agency
}
\DeclareAcronym{EMRI}{
    short = EMRI,
    long  = extreme-mass-ratio inspiral
}
\DeclareAcronym{ET}{
    short = ET,
    long  = Einstein Telescope
}
\DeclareAcronym{FPR}{
    short = FPR,
    long  = false positive rate
}
\DeclareAcronym{FAN}{
    short = FAN,
    long  = Fourier Analysis Network
}
\DeclareAcronym{GB}{
    short = GB,
    long  = galactic binary,
    long-plural-form = galactic binaries
}
\DeclareAcronym{GR}{
    short = GR,
    long  = general relativity
}
\DeclareAcronym{GW}{
    short = GW,
    long  = gravitational wave
}
\DeclareAcronym{LDC}{
    short = LDC,
    long  = LISA Data Challenge
}
\DeclareAcronym{LIGO}{
    short = LIGO,
    long  = \href{http://www.ligo.caltech.edu/}{Laser Interferemeter Gravitational Wave Observatory}
}
\DeclareAcronym{LISA}{
    short = LISA,
    long  = \href{https://www.lisamission.org/}{Laser Interferometer Space Antenna}
}

\DeclareAcronym{LSO}{
    short = LSO,
    long  = last stable orbit
}
\DeclareAcronym{MLP}{
    short = MLP,
    long  = multilayer perceptron
}

\DeclareAcronym{MBH}{
    short = MBH,
    long  = massive black hole
}
\DeclareAcronym{MBHB}{
    short = MBHB,
    long  = massive black hole binary,
    long-plural-form = massive black hole binaries
}

\DeclareAcronym{MCMC}{
    short = MCMC,
    long  = Markov-chain Monte Carlo
}
\DeclareAcronym{MLDC}{
    short = MLDC,
    long  = \href{http://astrogravs.nasa.gov/docs/mldc/}{Mock LISA Data Challenge}
}
\DeclareAcronym{NK}{
    short = NK,
    long  = numerical kludge
}
\DeclareAcronym{NSBH}{
    short = NS-BH,
    long  = neutron star-black hole
}
\DeclareAcronym{OMS}{
    short = OMS,
    long  = optical metrology system
}
\DeclareAcronym{PSD}{
    short = PSD,
    long  = power spectral density
}
\DeclareAcronym{ReLU}{
    short = ReLU,
    long  = rectified linear 
 unit
}

\DeclareAcronym{ROC}{
    short = ROC,
    long  = receiver operating characteristic
}
\DeclareAcronym{SGWB}{
    short = SGWB,
    long  = stochastic gravitational wave background
}
\DeclareAcronym{SMBH}{
    short = SMBH,
    long  = super-massive black hole
}
\DeclareAcronym{SNR}{
    short = SNR,
    long  = signal-to-noise ratio
}
\DeclareAcronym{SOBH}{
    short = SOBH,
    long  = stellar origin black hole binary
}
\DeclareAcronym{SSB}{
    short = SSB,
    long  = solar system barycenter
}
\DeclareAcronym{TCN}{
    short = TCN,
    long  = temporal convolutional network
}
\DeclareAcronym{TDI}{
    short = TDI,
    long  = time delay interferometry
}
\DeclareAcronym{TPR}{
    short = TPR,
    long  = true positive rate
}
\DeclareAcronym{t-SNE}{
    short = t-SNE,
    long  = t-distributed stochastic neighbor embedding
}
\DeclareAcronym{VGB}{
    short = VGB,
    long  = verification galactic binary,
    long-plural-form = verification galactic binaries
}

\begin{document}

\preprint{APS/123-QED}

% \title{Counting and Separating Compact Binary Coalescence \\ Gravitational Wave Signals Using Multi-Decoder Dual-Path FANFormer}
\title{Compact Binary Coalescence Gravitational Wave Signals Counting and Separation}

\author{Tianyu Zhao \orcidlink{0000-0003-3121-6042}}
\affiliation{Center for Gravitational Wave Experiment, National Microgravity Laboratory, Institute of Mechanics, Chinese Academy of Sciences, Beijing, 100190, China}

\author{Yue Zhou \orcidlink{0000-0003-2771-0144}}
\affiliation{Peng Cheng Laboratory, Shenzhen, 518055, China}

\author{Ruijun Shi \orcidlink{0009-0001-3320-8049}}%
\affiliation{School of Physics and Astronomy, Beijing Normal University, Beijing, 100875, China}
\affiliation{Institute for Frontiers in Astronomy and Astrophysics, Beijing Normal University, Beijing, 102206, China}

\author{\\Peng Xu \orcidlink{0000-0002-3543-7777}}
\thanks{Corresponding author: \href{mailto:xupeng@imech.ac.cn}{xupeng@imech.ac.cn}}
\affiliation{Center for Gravitational Wave Experiment, National Microgravity Laboratory, Institute of Mechanics, Chinese Academy of Sciences, Beijing, 100190, China}
\affiliation{Taiji Laboratory for Gravitational Wave Universe (Beijing/Hangzhou), University of Chinese Academy of Sciences (UCAS), Beijing, 100049, China}
\affiliation{Key Laboratory of Gravitational Wave Precision Measurement of Zhejiang Province, Hangzhou Institute for Advanced Study, UCAS, Hangzhou, 310024, China}
\affiliation{Lanzhou Center of Theoretical Physics, Lanzhou University, Lanzhou, 730000, China}

\author{Zhoujian Cao \orcidlink{0000-0002-1932-7295}}%
\thanks{Corresponding author: \href{mailto:zjcao@bnu.edu.cn}{zjcao@bnu.edu.cn}}
\affiliation{School of Physics and Astronomy, Beijing Normal University, Beijing, 100875, China}
\affiliation{Institute for Frontiers in Astronomy and Astrophysics, Beijing Normal University, Beijing, 102206, China}
\affiliation{School of Fundamental Physics and Mathematical Sciences, Hangzhou Institute for Advanced Study, UCAS, Hangzhou, 310024, China}

\author{Zhixiang Ren \orcidlink{0000-0002-4104-3790}}
\thanks{Corresponding author: \href{mailto:renzhx@pcl.ac.cn}{renzhx@pcl.ac.cn}}
\affiliation{Peng Cheng Laboratory, Shenzhen, 518055, China}

\date{\today}% It is always \today, today,

\begin{abstract}
	% 1. current issue
	As next-generation gravitational-wave (GW) observatories approach unprecedented sensitivities, the need for robust methods to analyze increasingly complex, overlapping signals becomes ever more pressing.
	% 2. existing method
	Existing matched-filtering approaches and deep-learning techniques can typically handle only one or two concurrent signals, offering limited adaptability to more varied and intricate superimposed waveforms.
	% 3. our method
	To overcome these constraints, we present the UnMixFormer, an attention-based architecture that not only identifies the unknown number of concurrent compact binary coalescence GW events but also disentangles their individual waveforms through a multi-decoder architecture, even when confronted with five overlapping signals.
	% 4. our innovation
	% By integrating a multi-decoder mechanism, 
	Our UnMixFormer is capable of capturing both short- and long-range dependencies by modeling them in a dual-path manner, while also enhancing periodic feature representation by incorporating Fourier Analysis Networks.
	Our approach adeptly processes binary black hole, binary neutron star, and neutron star–black hole systems over extended time series data (16,384 samples).
	% 5. our results
	When evaluating on synthetic data with signal-to-noise ratios (SNR) ranging from 10 to 50, our method achieves 99.89\% counting accuracy, a mean overlap of 0.9831 between separated waveforms and templates, and robust generalization ability to waveforms with spin precession, orbital eccentricity, and higher modes, marking a substantial advance in the precision and versatility of GW data analysis.

\end{abstract}

% \keywords{Suggested keywords}
% Use showkeys class option if keyword
% display desired

\maketitle

% reset acromyms' appearence
\acbarrier

% matched filtering-based and deep learning-based method
%  todo: interpret not t-sne, 
%  kernel visualization,
%  before after kernel

%  showcase input output different signals (science) 
%  3 performance, quantitatively
%  heatmap physical meaning appendix
\section{\label{sec:intro} Introduction}
% GW discovery and space-based GW detection

\Ac{GW} astronomy has experienced tremendous progress since the first detection of a binary black hole merger by \ac{LIGO} and Virgo in 2015 \cite{abbott_observation_2016}. The sensitivity of ground-based detectors has continuously improved, enabling the detection of an increasing number of astrophysical events, offering new insights into the nature of compact binaries \cite{cahillane_review_2022,martynov_sensitivity_2016}. The future of \ac{GW} detection is more promising with the introduction of third-generation detectors such as the \ac{ET} \cite{maggiore_science_2020}, \ac{CE} \cite{evans_horizon_2021} and space-based observatories like \ac{LISA} \cite{amaro-seoane_laser_2017}, Taiji \cite{hu_taiji_2017,ren_taiji_2023}, and TianQin \cite{luo_tianqin_2016} set to vastly expand the range and frequency of detectable signals. These advanced instruments will provide unprecedented opportunities to study a broad spectrum of sources, including stellar-mass binaries, massive black hole binaries, and a wide variety of \acp{CBS} \cite{bailes_gravitational-wave_2021}.

\begin{figure*}[ht!]
	\centering
	\includegraphics[width=0.7\textwidth]{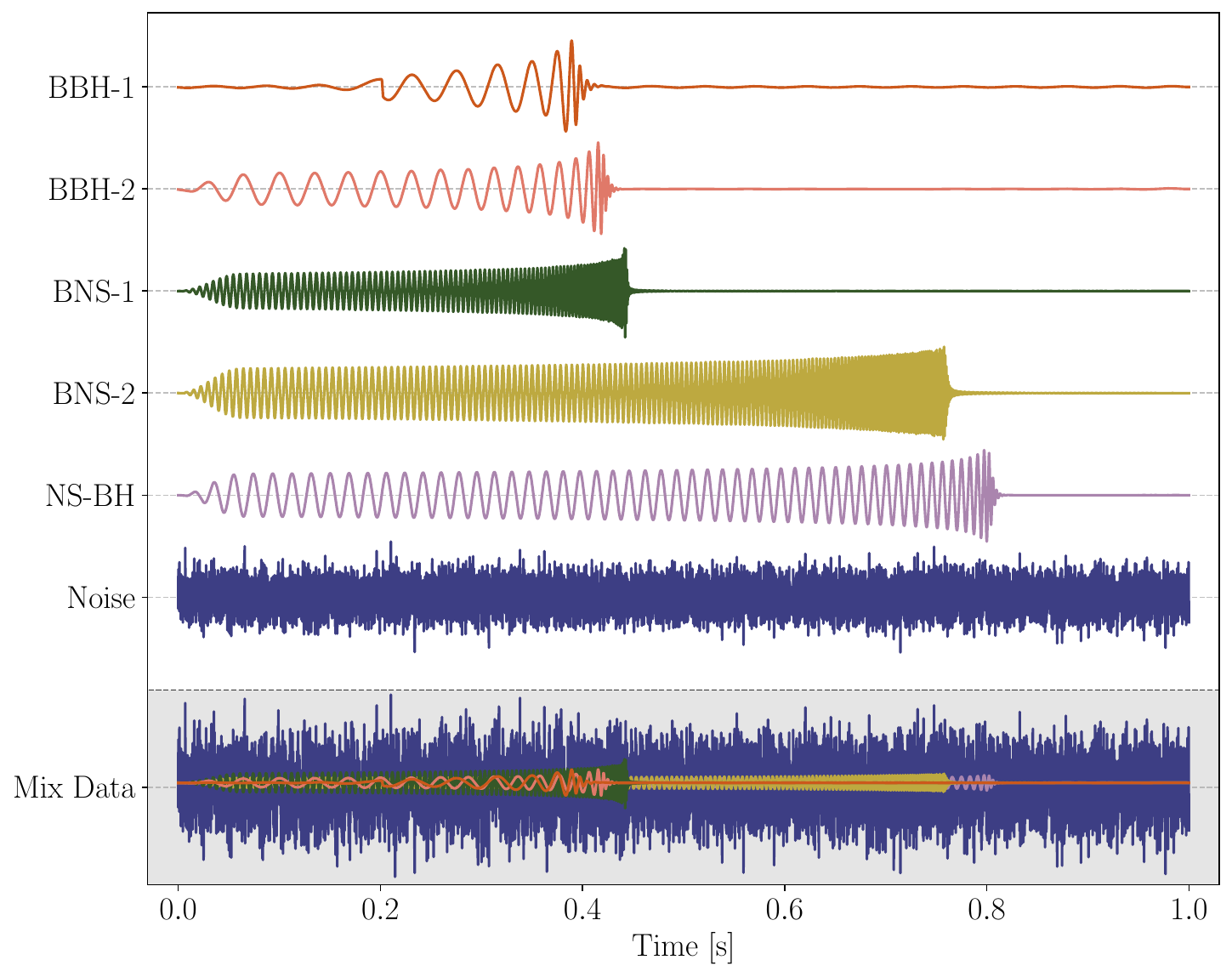}
	\caption{\textbf{Illustration of training data with overlapping signals.} This figure demonstrates the overlapping of multiple GW signals, including two BBH (BBH-1, BBH-2), two BNS (BNS-1, BNS-2), and one NS-BH. The individual signals are projected onto the detector with SNRs of 12, 15, 30, 20, and 15, respectively. The corresponding chirp masses are 45.14, 24.44, 1.63, 1.38, and 4.97 \(M_\odot\). The bottom waveform depicts the combined data, reflecting the realistic challenges of signal counting and separation in GW data analysis.
	}
	\label{fig:data}
\end{figure*}

While the improved sensitivity of next-generation detectors promises to revolutionize our understanding of the universe, it also presents significant challenges, particularly the increasing rate of overlapping signals from \acp{CBC} \cite{wang_rigorous_2025,iacovelli_forecasting_2022,pieroni_detectability_2022}. In space-based detectors, the overlap problem is complicated by distinct signal types such as \acp{MBHB}, \acp{EMRI}, and \acp{GB}, which can overlap both temporally and spectrally \cite{digman_lisa_2022,arca_sedda_merging_2021,baghi_2022}. However, the ground-based scenario introduces its own set of complexities \cite{himemoto_impacts_2021}. With third-generation detectors, the expected increase in event rate will result in more frequent overlaps from different types of \acp{CBC}, including \ac{BBH}, \ac{BNS}, and \ac{NSBH} mergers \cite{iacovelli_forecasting_2022}. Although these signals belong to the \ac{CBC} family, the increased event rate and the overlapping frequency range create significant difficulties in detection \cite{chamberlain_theoretical_2017}. Traditional matched filtering, which assumes isolated signals with well-defined templates, becomes increasingly ineffective in the case of multiple overlapping signals \cite{johnson_source_2024,hourihane_accurate_2022,regimbau_gravitational-wave_2009}. When events overlap, their parameter space expands dramatically, making accurate detection computationally costly \cite{wu_mock_2023,badaracco_blind_2024}.
Furthermore, parameter estimation becomes challenging as overlapping signals will introduce biases \cite{relton_parameter_2021,samajdar_biases_2021,pizzati_toward_2022}. If multiple signals are treated as a single event, the resulting parameter estimation, such as the mass, spin, and distance of the sources, can be significantly distorted \cite{janquart_analyses_2023,antonelli_noisy_2021}. This not only reduces the accuracy of individual measurements, but also complicates the correct identification and classification of events \cite{wang_anatomy_2024,relton_addressing_2022}. Moreover, while increased detector sensitivity can boost the overall detectability of a stochastic gravitational-wave background (SGWB) \cite{kerachian_detectability_2024}, the accompanying increase overlap rate also limiting the power to test general relativity under such conditions \cite{hu_accumulating_2023,dang_impact_2024}, highlighting the need for innovative data analysis methods that are capable of efficiently and accurately separating and identifying overlapping signals \cite{pieroni_detectability_2022}.

Recent advances in \ac{AI} have significantly improved multiple aspects of \ac{GW} data analysis \cite{zhao_dawn_2023}. \ac{AI}-based methodologies enable rapid generation of waveforms \cite{shi_compact_2024,shi_rapid_2025} and highly accurate detection of \ac{GW} signals \cite{george_deep_2018,gabbard_matching_2018} even under substantial noise \cite{zhao_dilated_2024, huerta_accelerated_2021}, and effectively extract waveforms from complex backgrounds \cite{zhao_space-based_2023,wei_gravitational_2020,wei_applications_2021,wang_waveformer_2024}.
These approaches have additionally demonstrated considerable promise in parameter estimation, reliably inferring fundamental source properties from noisy observational data \cite{dax_real-time_2021,dax_neural_2023,wildberger_adapting_2023}.
Despite these achievements, a longstanding challenge persists: robust separation of overlapping \ac{GW} signals.
Current \ac{AI} models are predominantly designed for \ac{BBH} systems and are commonly restricted to scenarios involving only two concurrent signals \cite{langendorff_normalizing_2023,alvey_what_2023}.
This limited scope poses pronounced difficulties when addressing more intricate cases, including overlaps of more than two signals or those arising from alternative compact binary sources, such as \ac{BNS} or \ac{NSBH} mergers. While certain methods have successfully disentangled BBH signals, they have demonstrated limited generalizability to more complex multi-signal scenarios \cite{langendorff_normalizing_2023,alvey_what_2023,ma_gravitational_2024}, underscoring the need for more versatile and resilient approaches.

\begin{figure*}[ht!]
	\centering
	\includegraphics[width=\textwidth]{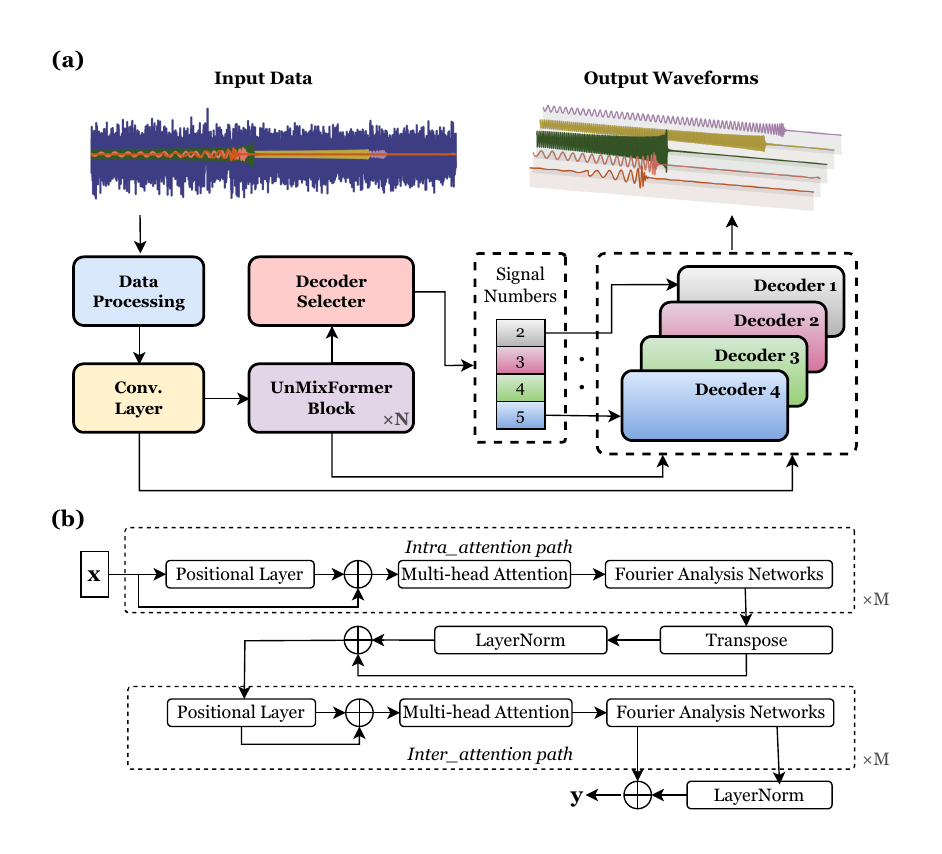}
	\caption{\textbf{UnMixFormer Architecture.}
		(a). The overall framework for counting and separating  overlapping GW signals. We firstly employ CNN-based encoders to extract data embeddings, which are then fused and passed into UnMixFormer blocks. The counting head predicts the number of sources and activates the appropriate decoder to reconstruct individual waveforms. (b). The core UnMixFormer block operates with intra- and inter-attention mechanisms to capture fine-grained local features and global context. FAN layers in the feedforward module enhance periodic feature modeling and the positional encoding incorporates sequential information, enabling efficient separation of overlapping signals.
	}
	\label{fig:model}
\end{figure*}

In this paper, we introduce the UnMixFormer, a novel model designed to accurately count and separate overlapping \ac{GW} signals originating from \ac{CBC} events, including \ac{BBH}, \ac{BNS}, and \ac{NSBH} mergers. The architecture employs a multi-decoder mechanism to determine the number of overlapping signals and utilizes a dual-path transformer framework to effectively and efficiently model both short-range and long-range dependencies within the data. Additionally, we incorporate a \ac{FAN} into the transformer block to enhance the reconstruction of periodic components that are particularly important for capturing high-frequency structures in signal waveforms.
Our model achieves a counting accuracy of 99.89\% and an average overlap of 0.9831 in signal separation, demonstrating high fidelity in reconstructing individual waveforms from overlapping signals. Notably, the model also generalizes well to more complex waveforms and accurately separates signals that exhibit eccentricity and precession. This advancement holds significant promise for improving the analysis of overlapping \ac{GW} signals, contributing to more precise astrophysical interpretations and discoveries.

The remainder of this paper is organized as follows: \Cref{sec:method} provides a comprehensive overview of the data generation process and delineates the architecture of our proposed model, the UnMixFormer. In \Cref{sec:result}, we present the outcomes of our extensive experiments on \ac{GW} signal counting and separation, highlighting the efficacy of our approach. Finally, \Cref{sec:discussion} concludes the paper with a summary of our key findings and outlines potential avenues for future research inspired by these results.

\section{\label{sec:method} Method}
% \subsection{EMRI Waveform Modeling}
% difficulties and importance of EMRI waveform modeling
% Existing EMRI template, their assumptions and difference
\begin{figure*}[!ht]
	\centering
	\subfloat[\label{fig:conf_mat}]{%
		\includegraphics[width=0.43\textwidth]{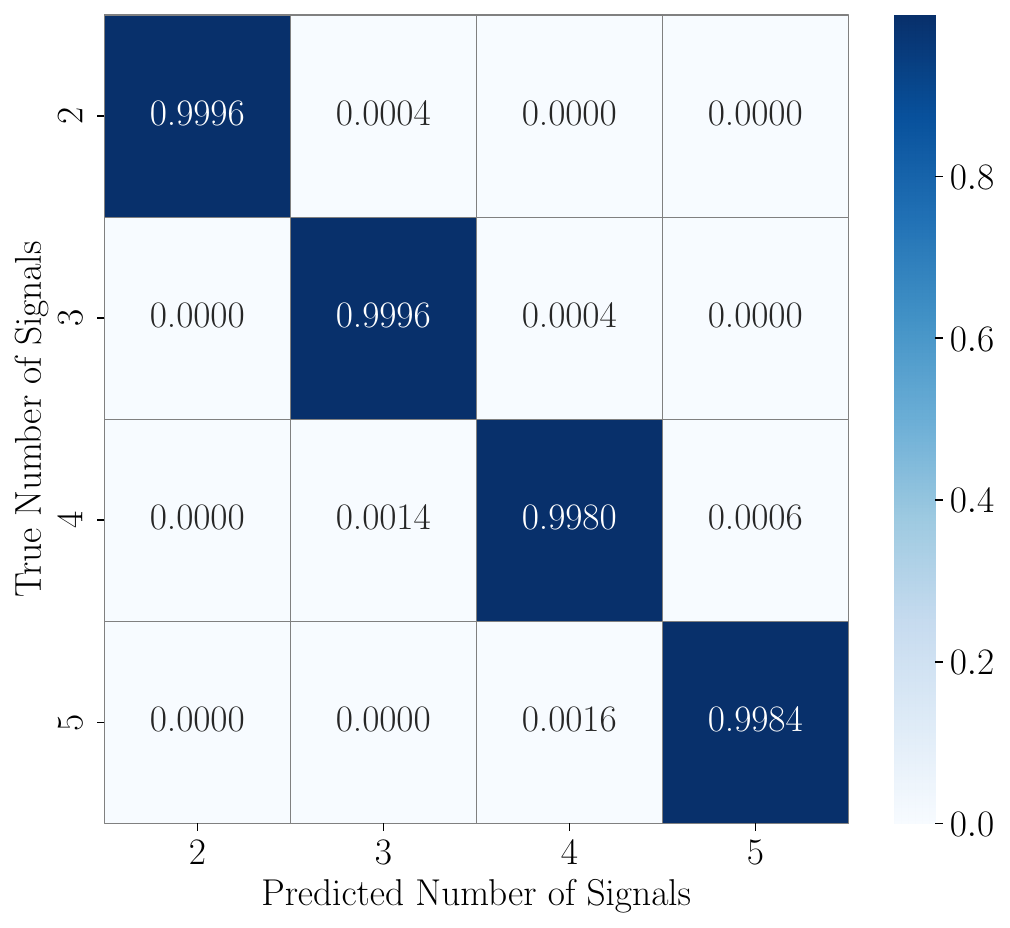}
	}\hfill
	\subfloat[\label{fig:roc}]{%
		\includegraphics[width=0.54\textwidth]{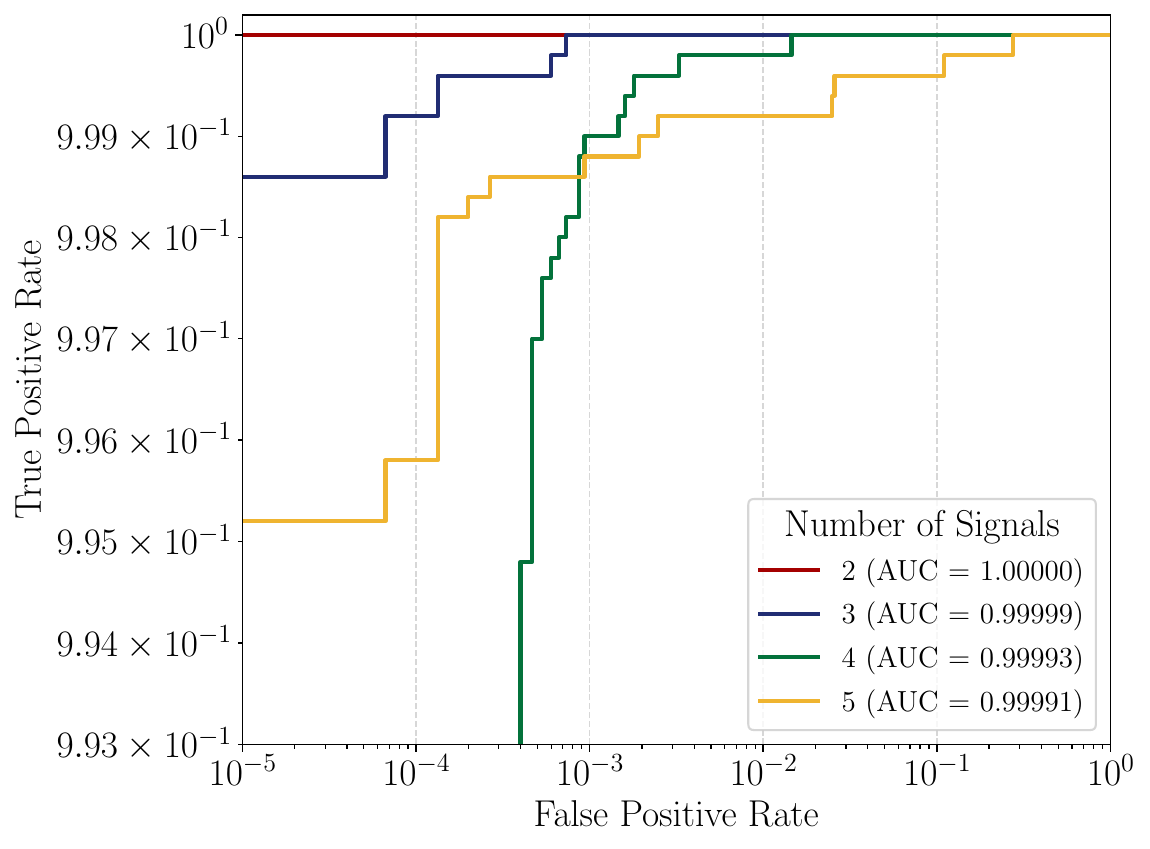}
	}
	% \hfill
	% \subfloat[\label{fig:acc_snr}]{%
	%     \includegraphics[width=0.32\textwidth]{acc_snr.pdf}
	% }
	\caption{\textbf{Counting performance of overlapping CBC signals.} (a). The normalized confusion matrix shows the high accuracy of predicting the number of overlapping signals, with correct predictions dominating the diagonal entries (2 to 5 signals). (b). ROC curves illustrating the performance of signal counting for varying numbers of signals. The curves demonstrate near-perfect detection across all cases.
	}
	\label{fig:count}
\end{figure*}

\subsection{Data Curation}
The dataset used in this study was generated using the \texttt{PyCBC} \cite{dal_canton_real-time_2021} library, which provides a comprehensive suite of tools for simulating \ac{GW} signals and noise. Each data sample spans one second in duration and is sampled at 16,384 Hz, reflecting the high-fidelity demands of next-generation detectors. The noise simulation relied on the \ac{PSD} of the \ac{CE}-40km detector to produce Gaussian noise. This noise was then superimposed onto the \ac{GW} signals, yielding realistic observational data that faithfully approximate the detector’s response to both signal and noise. Under these conditions, the observed time-domain data \( d(t) \) can be expressed as:

\begin{equation}
	d(t) = \sum_{i=1}^{K} s_i(t) + n(t),
\end{equation}
where \( K \) is the number of signals present (ranging from 2 to 5 signals per sample). An example of such a data sample is illustrated in \Cref{fig:data}. The signals \( s_i(t) \) are generated using different waveform templates and sampled parameters as described below.

\paragraph{Signal waveform}
The \ac{GW} signals are simulated for three types of sources. The first step is to assign a class label to each event, selecting one of the labels: \texttt{["BBH", "NS-BH", "BNS"]}. After selecting the label, the parameters for the signal are sampled from predefined priors.

For each type of signal, the component masses, \( m_{\mathrm{BH}} \) for black holes and \( m_{\mathrm{NS}} \) for neutron stars, are sampled uniformly within their respective ranges. The spin parameters \( s_1^z \) and \( s_2^z \), coalescence phase \( \phi_c \), inclination angle \( \iota \), polarization angle \( \psi \), right ascension \( \alpha \), declination \( \delta \), and coalescence time \( t_c \) are  uniformly drawn from their respective prior distributions. A summary of the parameter ranges can be found in \Cref{tab:priors}. For \ac{BBH}, the \texttt{SEOBNRv4} \cite{bohe_improved_2017} waveform template is used; for \ac{NSBH} systems, the \texttt{IMRPhenomT} \cite{estelles_new_2022} waveform template is employed; and for \ac{BNS}, the \texttt{TaylorF2} \cite{messina_quasi-55pn_2019} waveform template is utilized. These parameters are then used to generate the corresponding waveforms for each source.

\paragraph{Detector response}
After simulating the \ac{GW} signal, we obtain the two polarization states, \( h_+(t) \) and \( h_\times(t) \). These components are then projected onto the detector based on its orientation and the location of the source. The detector response to the signal is determined by the antenna pattern functions \( F_+( \alpha, \delta, \psi) \) and \( F_\times( \alpha, \delta, \psi) \), which depends on the sky location of the source, as well as the polarization angle \( \psi \). The detailed expressions for these antenna pattern functions can be found in \cite{whelan_visualization_2012}. The full signal observed by the detector, \( s_i(t) \), can be expressed as:
\begin{equation}
	s_i(t) = F_+( \alpha, \delta, \psi) h_+(t) + F_\times( \alpha, \delta, \psi) h_\times(t),
\end{equation}
To simulate the signal at the detector, we rescale each signal by its optimal \ac{SNR}. The optimal \ac{SNR} is calculated based on the inner product between the signal \( s(t) \) and the template \( h(t) \) in the frequency domain. The inner product is defined as:
\begin{equation}
	(a \mid b) = 2\int_{f_{\mathrm{min}}}^{f_{\mathrm{max}}} \frac{\tilde{a}^*(f)\tilde{b}(f)+\tilde{a}(f)\tilde{b}^*(f)}{S_n(f)}\, \dd{f} \qs
\end{equation}
where $\tilde{a}(f)$ and $\tilde{b}(f)$ represent the frequency domain signals, and the superscript $*$ denotes the complex conjugate. $S_n(f)$ is the one-side noise \ac{PSD}. The optimal \ac{SNR} is then calculated as:
\begin{equation}
	\text{SNR}_i^2 = (s_i \mid s_i) \qc
\end{equation}
then the signal is rescaled accordingly to match the desired optimal SNR.
In addition, we calculate the overlap between two signals as a measure of their shape similarity. The overlap \( \mathcal{O} \) between two waveform \( h_1(t) \) and \( h_2(t) \) is defined as the normalized inner product:
\begin{equation}
	\mathcal{O}(h_1,h_2) = \frac{(h_1 | h_2)}{\sqrt{(h_1 | h_1)(h_2 | h_2)}},
\end{equation}
and the mismatch $\mathcal{M}$ is:
\begin{equation}
	\mathcal{M}(h_1, h_2) = 1 - \mathcal{O}(h_1,h_2) .
\end{equation}
These metrics are useful in evaluating our model's performance.

\begin{figure*}[ht!]
	\centering
	\includegraphics[width=\textwidth]{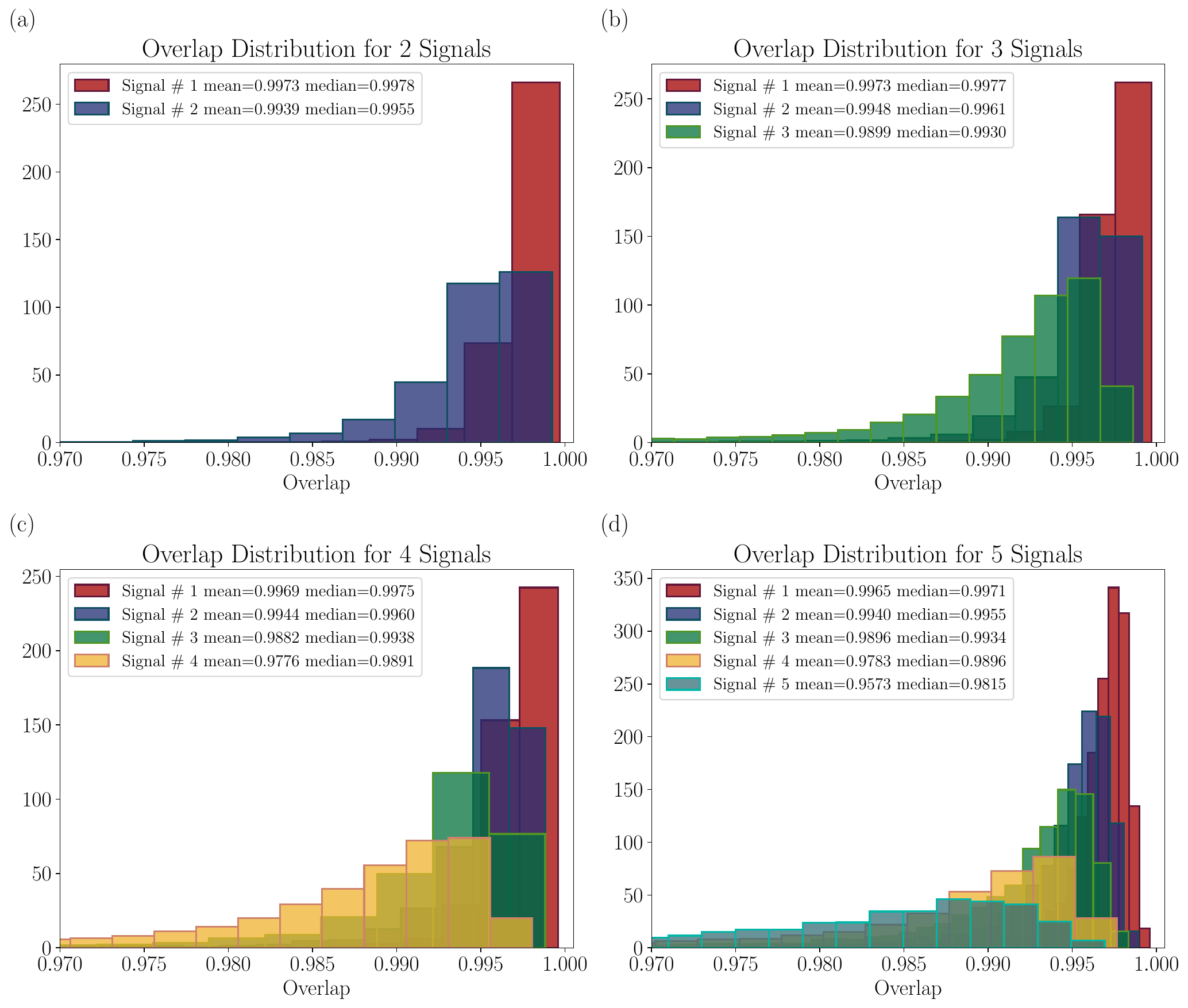}
	\caption{\textbf{Overlap distribution for separated signals.} Histograms of the overlap between separated waveforms and their corresponding target templates for cases with (a) 2 signals, (b) 3 signals, (c) 4 signals, and (d) 5 signals. The overlap measures the similarity between the separated and ground-truth signals, with values close to 1 indicating high accuracy. Each color shows the distributions for individual signals (Signal 1, Signal 2, etc.) across the dataset, demonstrating consistent separation quality as the number of overlapping signals increases.}
	\label{fig:hist}
\end{figure*}

\paragraph{Dataset composition}
The data is whitened before being used for training and testing. During the whitening process, a Tukey window with a parameter of \( \alpha = 1/8 \) is applied to minimize edge effects and reduce spectral leakage. After preprocessing, the dataset is split into training and testing set. The training set consists of 25,000 samples for each signal number case (with 2 to 5 signals), resulting in a total of 100,000 training samples. The testing set contains 5,000 samples per signal number case, with a total of 20,000 testing samples.

\begin{table}[htbp]
	\begin{center}
		\caption{Summary of parameter priors used in \ac{GW} signal simulation.}\label{tab:priors}
		\begin{tabular}{@{}lcc@{}}
			\toprule
			\hline
			\textbf{Parameter} & \textbf{Lower bound} & \textbf{Upper bound} \\
			\hline\hline
			$m_{\mathrm{BH}}$  & $10$                 & $75$                 \\
			$m_{\mathrm{NS}}$  & $1$                  & $3$                  \\
			$s_1^z, s_2^z$     & $-0.99$              & $0.99$               \\
			$\phi_c$           & $0$                  & $2\pi$               \\
			SNR                & $10$                 & $50$                 \\
			$\iota$            & $0$                  & $\pi$                \\
			$\psi$             & $0$                  & $\pi$                \\
			$\alpha$           & $0$                  & $2\pi$               \\
			$\delta$           & $0$                  & $\pi$                \\
			$t_c$              & $0.25$               & $0.75$               \\
			\hline
			\bottomrule
		\end{tabular}
	\end{center}
\end{table}

\subsection{UnMixFormer}

Our proposed UnMixFormer model directly addresses the challenge of accurately separating and counting overlapping \ac{GW} signals.An overview of the model architecture is shown in \Cref{fig:model}. Building upon a dual-path architecture \cite{chen_dual-path_2020,luo_dual-path_2020}, it integrates a fully attention-based backbone and a multi-decoder structure to jointly identify the number of sources and separate their individual waveforms. Although inspired by the multi-decoder DPRNN \cite{junzhe_multi-decoder_2020}, this framework introduces substantial enhancements that are capable of handling increasingly complex and diverse signal scenarios, and ultimately offer a more robust and flexible solution than existing methods.

\paragraph{Transformer encoder}
The backbone of the model is built on a transformer encoder architecture, which replaces traditional recurrent modules. Transformer encoders utilize a multi-head attention mechanism to effectively capture long-range dependencies across the input sequence. Each attention head operates using the query (\(Q\)), key (\(K\)), and value (\(V\)) formulation, defined as:
\[
	\text{Attention}(Q, K, V) = \text{softmax}\left(\frac{QK^T}{\sqrt{d_k}}\right)V,
\]
where \(d_k\) represents the dimension of the key vector. The multi-head attention mechanism combines the outputs from multiple attention heads, allowing the model to capture diverse dependencies. To incorporate positional information in the time-frequency domain, the transformer employs positional encoding, which enables it to process sequential data effectively. Additionally, each transformer layer includes a feedforward neural network for nonlinear transformations. This design enables the model to process both local and global dependencies within the dual-path segmentation structure, with the intra-segment path capturing fine-grained local patterns and the inter-segment path aggregating global context. The parallel processing capability of transformers significantly improves computational efficiency and enhances performance on overlapping signals.

\paragraph{FAN}
A key element of our model architecture is the integration of the \ac{FAN} \cite{dong_fan_2024} within the transformer blocks, replacing conventional \ac{MLP} layers with a FAN that capable of explicitly modeling periodicity. Unlike \acp{MLP}, which directly transform inputs through linear projections followed by nonlinear activations, \ac{FAN} leverages the principles of Fourier transform to decompose signals into periodic components. Formally, if \( x \in \mathbb{R}^{d_x} \) is an input and \( \phi(x) \) denotes a \ac{FAN} layer, we introduce learnable parameters \( W_p \) and \( W_{\bar{p}} \) that transform \( x \) into both cosine and sine terms as well as a nonlinear projection:
\[
	\phi(x) = [\cos(W_p x) \,||\, \sin(W_p x) \,||\, \sigma(B_{\bar{p}} + W_{\bar{p}} x)],
\]
where \( \sigma(\cdot) \) is a nonlinear activation. This design stands in contrast to a \ac{MLP} layer \(\Phi(x)\) that defined as
\[
	\Phi(x) = \sigma(B_{m} + W_{m} x),
\]
which relies solely on learned affine transformations and nonlinearities without explicit frequency modeling. By incorporating trigonometric functions into the transformation, FAN naturally encodes periodic features, thereby improving the model’s ability to disentangle overlapping signals with periodic structures. Additionally, FAN requires fewer parameters and computations than an MLP of comparable capacity, enabling a more efficient representation of the underlying signal periodicity.

\paragraph{Multi-decoder and selector}
To address the challenge of separating an unknown number of overlapping \ac{GW} signals, we assume a known upper limit \(K\) on the number of concurrent sources. The model first employs a counting head to estimate the number of signals, utilizing a classification approach that outputs a probability distribution over \(\{1, 2, \dots, K\}\). Given the ground-truth set of signals \(\mathbf{y}\), with \(|\mathbf{y}|\) denoting its cardinality, we define the cross-entropy loss as:
\begin{equation}
	\mathcal{L}_{\text{CE}}(\mathbf{x}, \mathbf{y}) = -\sum_{k=1}^K 1_{\{|\mathbf{y}|=k\}} \log \hat{p}(|\mathbf{y}|=k \mid \mathbf{x}),
\end{equation}
where \(\hat{p}(|\mathbf{y}|=k \mid \mathbf{x})\) is the predicted probability of input \(\mathbf{x}\) containing \(k\) signals, and \(1_{\{|\mathbf{y}|=k\}}\) is an indicator function that equals 1 if \(|\mathbf{y}|=k\) and 0 otherwise. Once the count is determined, the model activates the corresponding decoder-head that designed to reconstruct that number of sources. By dynamically selecting the appropriate decoder-head, a single model can flexibly handle variations in signal count, ensuring scalability across a range of overlapping signal scenarios.

\paragraph{Loss function}
After selecting the appropriate decoder-head, we optimize a separation objective to ensure high-fidelity waveform reconstruction. We adopt the permutation-invariant scale-invariant signal-to-noise ratio (SI-SNR) criterion \cite{vincent_performance_2006}. Given a ground-truth signal \(\mathbf{y}\) and a predicted signal \(\hat{\mathbf{y}}\), we first compute:
\begin{equation}
	\mathbf{y}_{\text{target}} = \frac{\langle \hat{\mathbf{y}}, \mathbf{y} \rangle}{\|\mathbf{y}\|^2}\mathbf{y}, \quad \mathbf{e} = \hat{\mathbf{y}} - \mathbf{y}_{\text{target}},
\end{equation}
and then define:
\begin{equation}
	\text{SI-SNR}(\mathbf{y}, \hat{\mathbf{y}}) = 20 \log_{10}\frac{\|\mathbf{y}_{\text{target}}\|}{\|\mathbf{e}\|}.
\end{equation}

For multiple sources, we select the permutation that maximizes the total SI-SNR, ensuring the best alignment between predictions and ground truth. The final loss balances counting accuracy and source reconstruction fidelity:
\begin{equation}
	\mathcal{L} = \lambda \cdot \mathcal{L}_{\text{SI-SNR}} + (1 - \lambda) \cdot \mathcal{L}_{\text{CE}},
\end{equation}
with \(\lambda\) controlling the trade-off. By jointly optimizing both terms, the model learns to reliably count sources while achieving high-quality waveform separation.

\subsection{Implementation Details}
The simulation of \ac{GW} signals and their embedding in realistic detector noise was conducted using the \texttt{PyCBC}\footnote{\url{https://pycbc.org}} library \cite{dal_canton_real-time_2021}. To facilitate robust and efficient time-series modeling, we used the \texttt{asteroid}\footnote{\url{https://github.com/asteroid-team/asteroid}} \cite{pariente_asteroid_2020}, \texttt{Time-Series-Library}\footnote{\url{https://github.com/thuml/Time-Series-Library}} \cite{wu_timesnet_2022,wang_deep_2024} and incorporated a FAN-based\footnote{\url{https://github.com/YihongDong/FAN}} architecture into our transformer blocks. The overall model was implemented in \texttt{PyTorch}, using the \texttt{Adam} \cite{kingma_adam_2015} optimizer with a learning rate of \(10^{-3}\) and no weight decay. Each dual-path block employed a chunk size of 64 and a hidden size of 256. We used 8 attention heads, two layers of Inter and intra transformer blocks. Early stopping based on validation performance and gradient clipping of 5 are employed. Training was performed on an NVIDIA GeForce RTX 4090 GPU with a batch size of 16 and lasted up to 100 epochs, which took about 47 hours. During inference, the model processes each batch in approximately 0.0015 seconds, yielding an efficient and flexible framework for separating complex \ac{GW} signals.

\section{\label{sec:result} Results}

\subsection{Signal Counting Accuracy}
Counting accuracy measures how well the predicted number of signals matches the true number in the testing set, defined as the ratio of correct predictions to the number of total samples. For multi-class problems, a confusion matrix is essential, with rows representing true counts and columns representing predictions. Diagonal entries indicate correct predictions while off-diagonal ones show errors. This analysis helps uncover error patterns, such as overestimation or underestimation, and provides insights into the model's performance across different classes.

The confusion matrix in \Cref{fig:conf_mat} reveals the model's strong capability to accurately count signals across all scenarios with the majority of predictions lying along the diagonal. For cases with two to five signals, misclassifications predominantly occur within a margin of ±1 signal, demonstrating the model's robustness in scenarios with varying levels of complexity. For instance, when the true number of signals is five, the model rarely misclassifies beyond four signals. Furthermore, the model achieves 100\% counting accuracy within a margin of ±1 signal across all test cases, underscoring its reliability in handling high-overlap scenarios and noisy environments.

% \begin{table}[ht!]
% \begin{center}
% \caption{\textbf{Counting Accuracy Over SNR}}\label{tab:sep_mean}%
% \begin{tabular}{c|ccccccccc}
% \toprule
% \hline
%  \textbf{No. of}  & \multicolumn{8}{c}{\textbf{SNR}} \\
% \cline{2-10}
% \textbf{Sig}  & 10 & 15 & 20 & 25 & 30 & 35 & 40 & 45 & 50 \\ \cline{1-10}
%  2 & 1.00 & 1.00 & 1.00 & 1.00 & 1.00 & 1.00 & 1.00 & 1.00 & 1.00 \\ 
%  3 & 0.75 & 0.98 & 1.00 & 1.00 & 1.00 & 1.00 & 1.00 & 1.00 & 1.00 \\
%  4 & 0.67 & 0.97 & 0.98 & 1.00 & 1.00 & 1.00 & 1.00 & 1.00 & 1.00 \\
%  5 & 0.60 & 1.00 & 1.00 & 1.00 & 1.00 & 1.00 & 1.00 & 1.00 & 1.00 \\
% \hline
%     \bottomrule
% \end{tabular}
% \end{center}
% \end{table}

\Cref{fig:roc} illustrates the Receiver Operating Characteristic (ROC) curves generated for varying numbers of overlapping signals (2, 3, 4, and 5). The ROC curves highlight the trade-off between the \ac{TPR} and \ac{FPR} across a wide range of thresholds. Our model consistently achieved near-perfect counting, as evidenced by the \ac{AUC} values of 1.00000, 0.99999, 0.99993, and 0.99991 for 2, 3, 4, and 5 signals, respectively. These results demonstrate the robustness of the model in distinguishing overlapping signals with minimal false positives, even in challenging scenarios with multiple sources. The high \ac{AUC} values confirm the efficacy of our approach in accurately identifying compact binary coalescence signals while maintaining reliability across diverse configurations.

\begin{figure*}[ht!]
	\centering
	\includegraphics[width=0.9\textwidth]{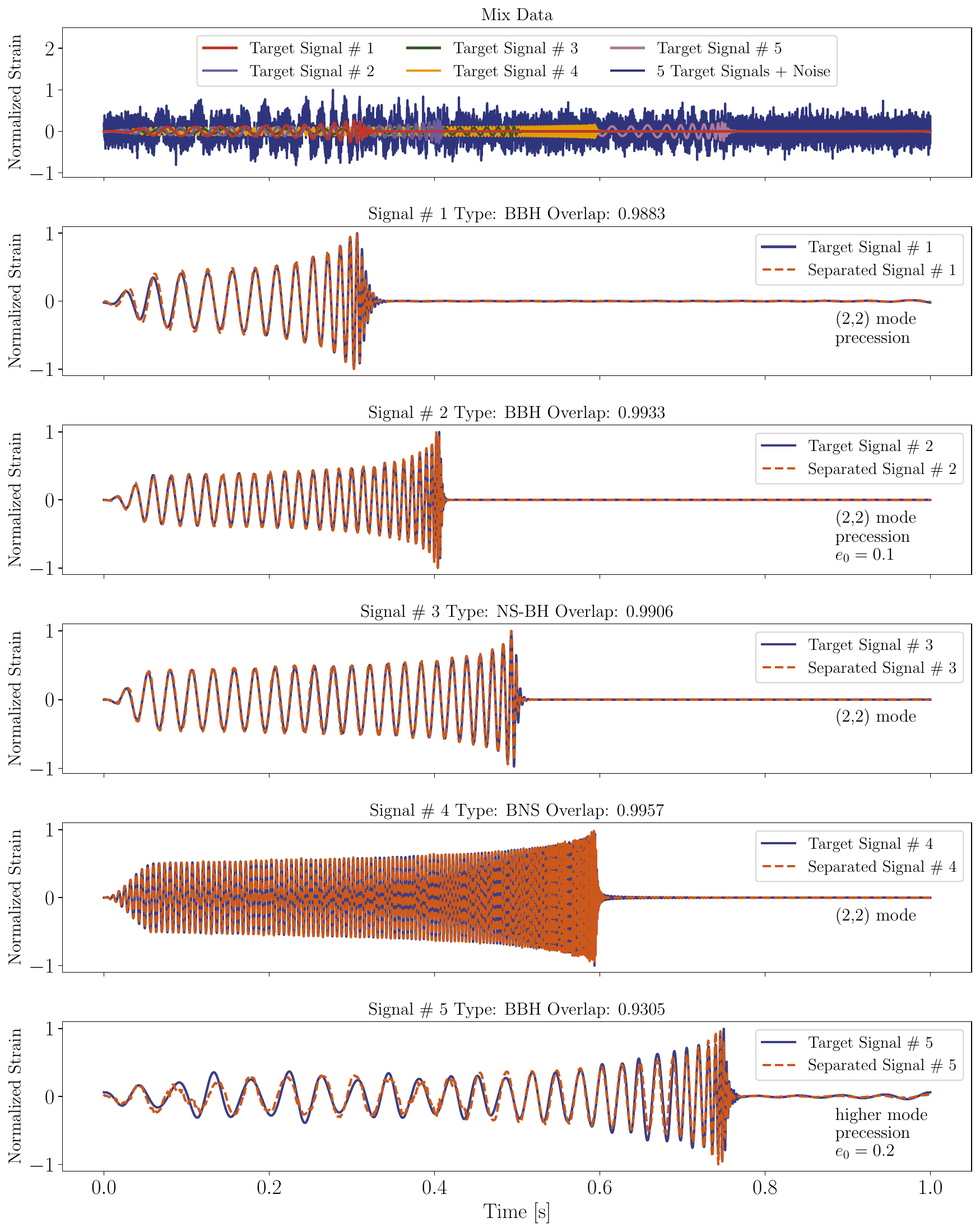}
	\caption{\textbf{Showcase of signal separation and generalization ability.} The top panel illustrates the mixed data that contains five target signals burried in noise. Other panels display the separated individual signals and target templates, including BBH waveforms with precession, orbital eccentricity, and higher modes, as well as NS-BH and BNS waveforms. The overlaps between the separated and target signals are consistently high, demonstrating the model's effectiveness in separating different types of signals and generalizing to diverse complex waveforms.
	}
	\label{fig:generalize}
\end{figure*}

\begin{figure*}[ht!]
	\centering
	\includegraphics[width=0.9\textwidth]{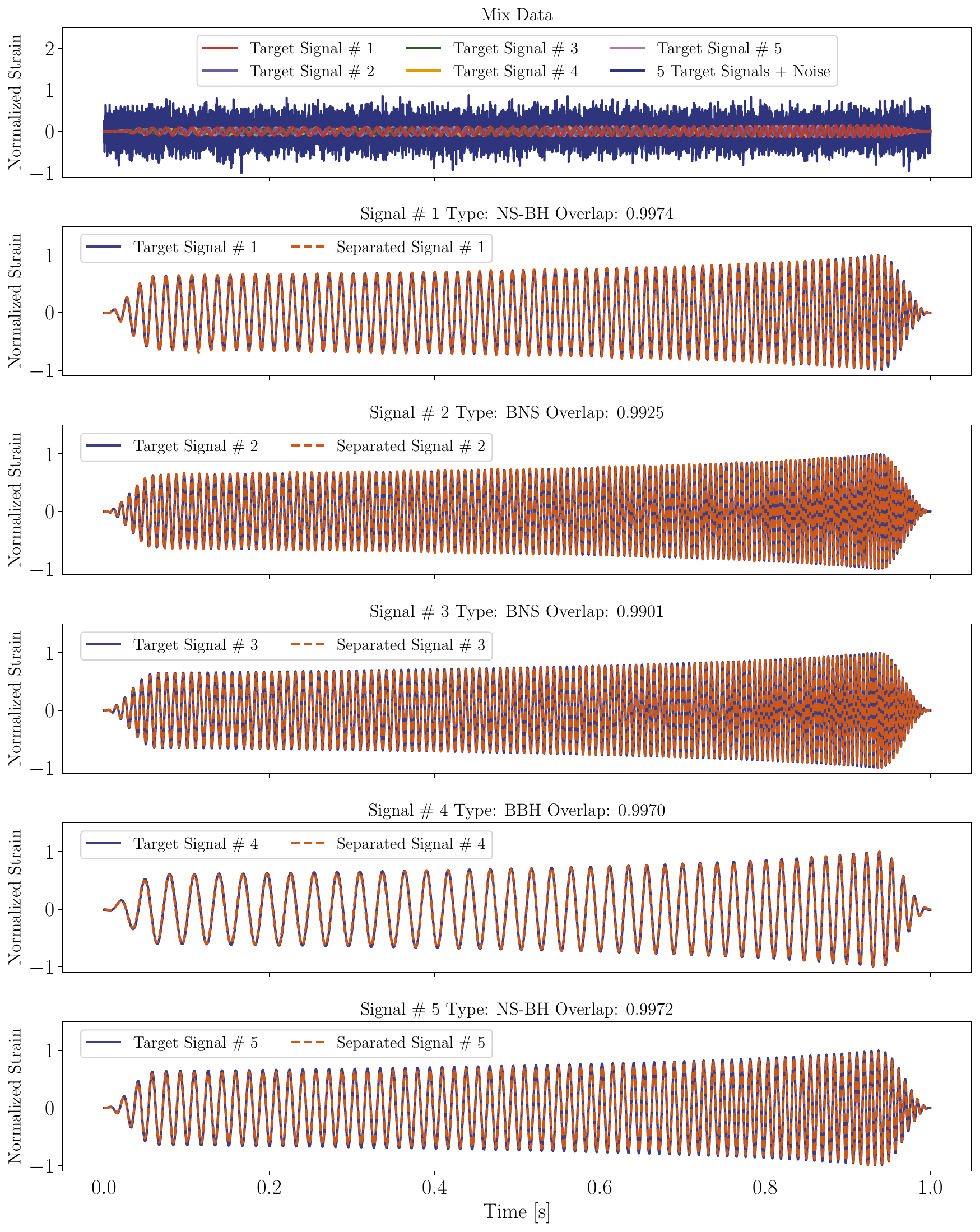}
	\caption{
		\textbf{Showcase of inspiral-only signal separation performance.} The top panel depicts mixed data containing five target inspiral-only signals buried in noise. The subsequent panels display the separated individual signals alongside their corresponding target templates, the overlap between separated signal and target templates is on the top of each panel, highlighting its robustness in scenarios relevant to third-generation detectors.
	}
	\label{fig:inspiral}
\end{figure*}

\subsection{Signal Separation Performance}
The separation performance of our model was first evaluated on the testing set, and the results are summarized in \Cref{fig:hist}. The histogram illustrates the distribution of overlap scores between the separated waveforms and their target templates. For cases involving multiple signals, the overlaps were sorted in descending order. This sorting ensures a clear visualization of the model's performance across different signals. By sorting, we highlight how the model achieves consistently higher overlaps for the most prominent signals, with secondary signals showing slightly reduced overlaps due to the increased complexity of separation. This approach provides a more transparent way to assess the model's separation capabilities. The mean and median overlap values for different numbers of signals are detailed in the upper-left corner of \Cref{fig:hist}, further affirming the model's robustness across varied scenarios.

\begin{figure}[ht!]
	\centering
	\includegraphics[width=0.47\textwidth]{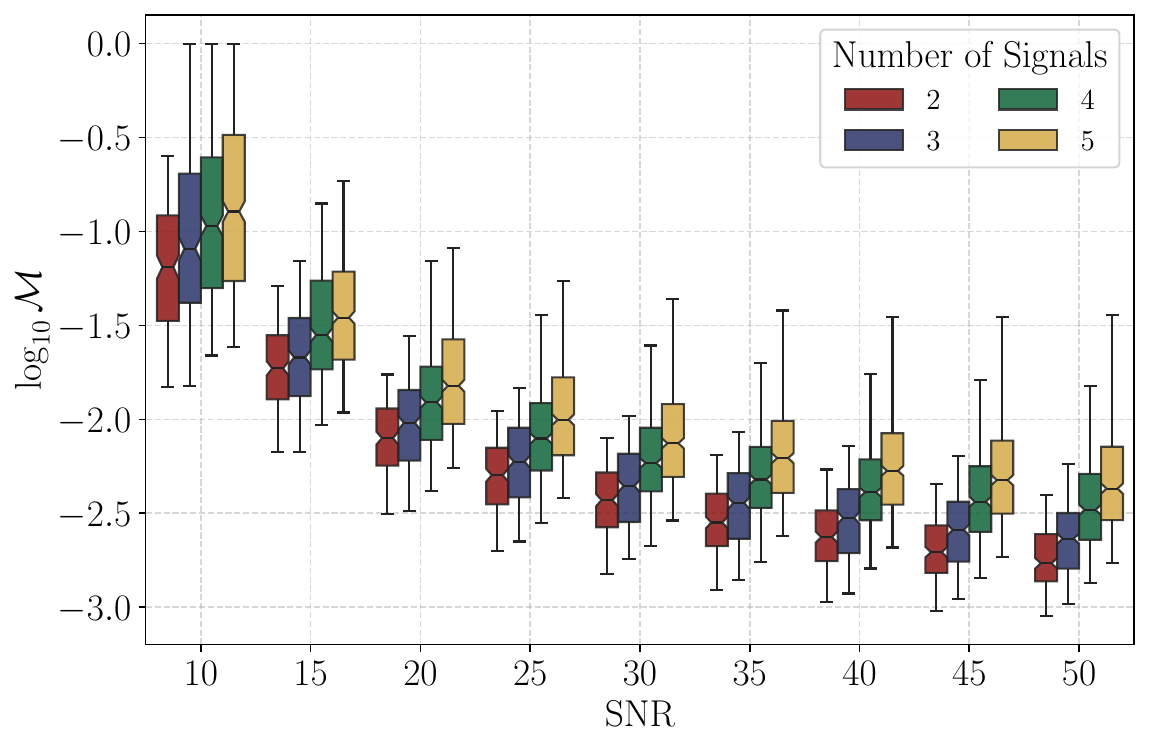}
	\caption{\textbf{Mismatch distributions of different SNR.} Boxplots illustrate the \(\log_{10} \mathcal{M}\)  values for separated waveforms across varying numbers of overlapping signals (2 to 5) and different SNR levels. The results demonstrate consistent performance with low mismatch values, even as the number of overlapping signals increases. Higher SNR levels correspond to better separation quality, emphasizing the model's robustness in challenging scenarios.
	}
	\label{fig:snr_mismatch}
\end{figure}

To further evaluate performance across different \ac{SNR} levels, we analyzed the separation accuracy as a function of SNR. As shown in \Cref{fig:snr_mismatch}, the results reveal that higher \ac{SNR} leads to significantly better separation performance, with overlaps exceeding 0.99 for \ac{SNR} above 20. In the case where the \ac{SNR} is equal to 10, some samples exhibit an overlap of 0. This occurs because the estimated number of signals exceeds the true number of signals, leaving no corresponding target for the extra estimated signal, resulting in an overlap of 0.
As the number of signals increases, the overlap decreases slightly, reflecting the inherent challenge of disentangling multiple overlapping waveforms. To better illustrate the separation performance across challenging cases, we plotted the log-transformed mismatch values, \(\log_{10} \mathcal{M}\), because the model achieves exceptional performance, with a significant proportion of test samples having overlaps exceeding 0.99. As the number of signals increases, the overlap decreases slightly, reflecting the inherent challenge of disentangling multiple overlapping waveforms. This analysis underscores the model's adaptability and reliability across varying signal conditions.

% \begin{table}[ht!]
% \begin{center}
% \caption{\textbf{Mean Overlap between separated waveforms and targets.}  
% This table presents the mean overlap between the separated waveforms and their corresponding targets for different numbers of overlapping signals (2–5). The results demonstrate the model's ability to achieve high fidelity in reconstructing individual waveforms, even as the number of overlapping signals increases.
% }\label{tab:sep_mean}%
% \begin{tabular}{@{}cccccc@{}}
%     \toprule
%     \hline
%     \textbf{No. of Sig} & \textbf{Sig 1} & \textbf{Sig 2} & \textbf{Sig 3} & \textbf{Sig 4} & \textbf{Sig 5} \\
%     \hline\hline
%     2 & 0.9973 & 0.9939 &  &  &  \\
%     3 & 0.9973 & 0.9948 & 0.9899 &  &  \\
%     4 & 0.9969 & 0.9944 & 0.9882 & 0.9776 &  \\
%     5 & 0.9965 & 0.9940 & 0.9896 & 0.9783 & 0.9573 \\
%     \hline
%     \bottomrule
% \end{tabular}
% \end{center}
% \end{table}

\begin{figure}[ht!]
	\centering
	\includegraphics[width=0.47\textwidth]{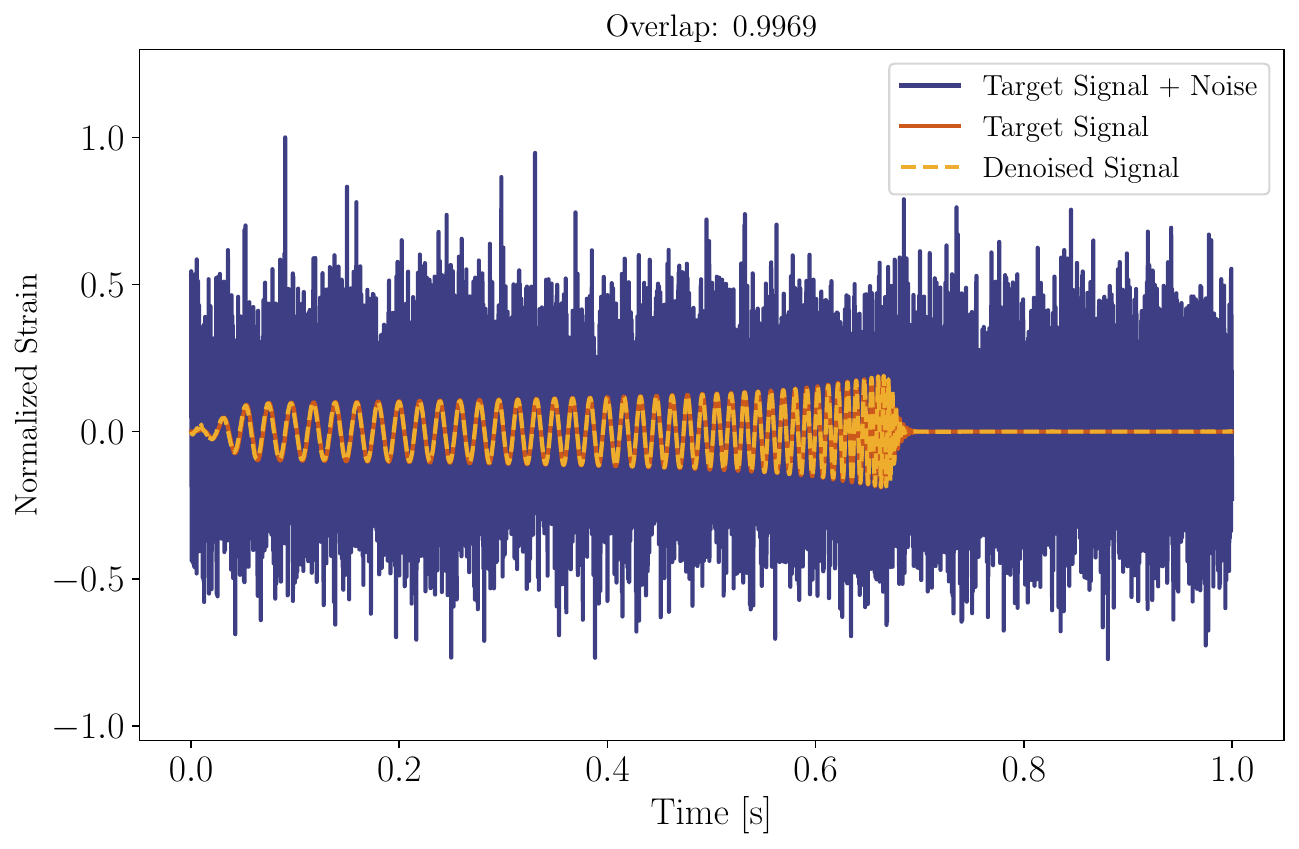}
	\caption{
		\textbf{Generalization ability to single signal denoising scenario.} This figure illustrates our model's capacity to denoise and reconstruct a single GW signal from noisy data. The high overlap demonstrating robust generalization ability to denoising scenarios.}
	\label{fig:denoise}
\end{figure}

% \subsection{Interpretability}

\begin{figure*}[htb!]
	\centering
	\subfloat[\label{fig:p_mismatch}]{%
		\includegraphics[width=0.49\textwidth]{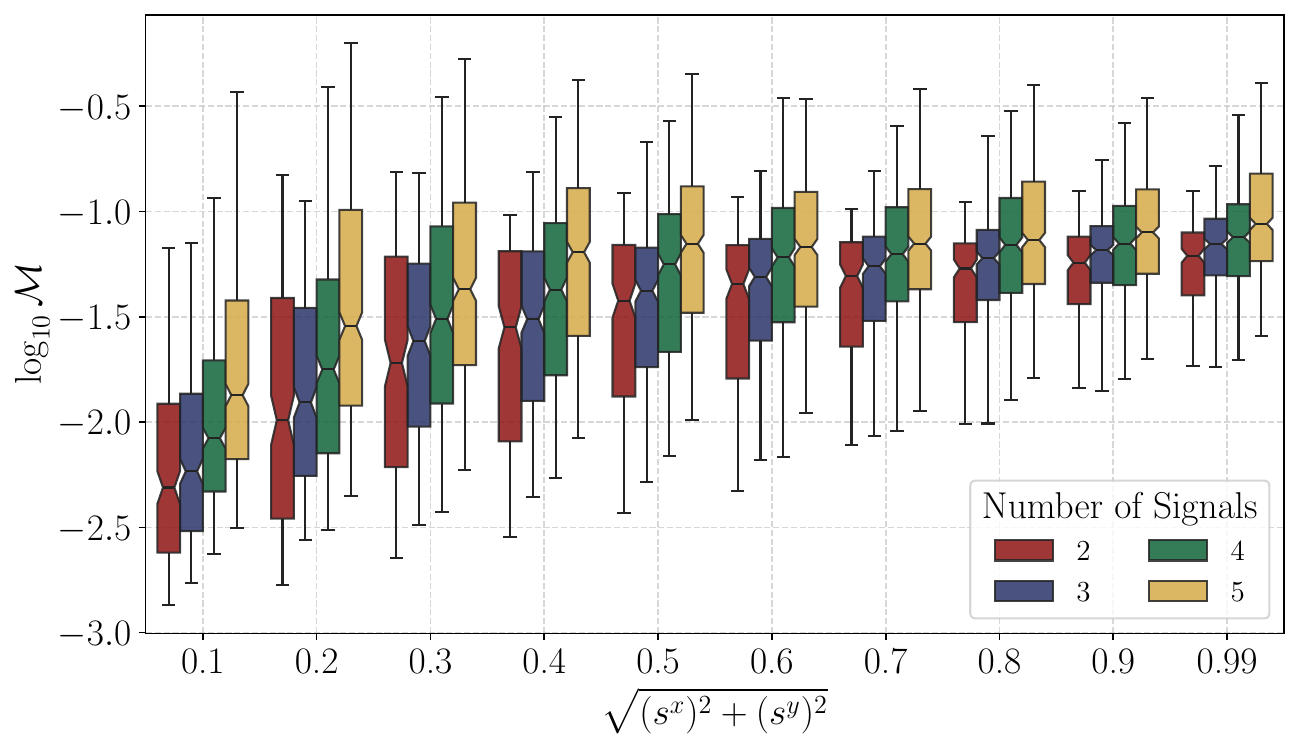}
	}\hfill
	\subfloat[\label{fig:e0_mismatch}]{%
		\includegraphics[width=0.49\textwidth]{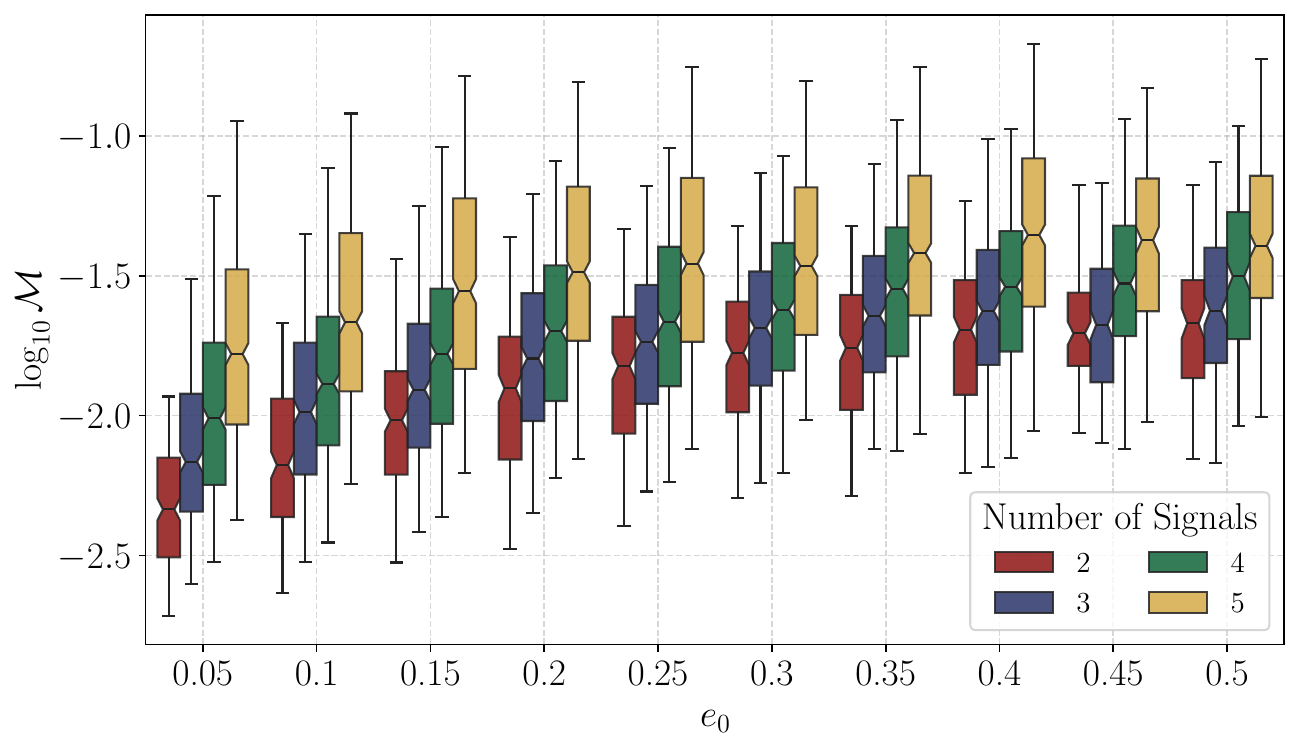}
	}
	\caption{\textbf{Generalization performance across different precession and orbital eccentricity.} (a) The mismatch (\( \log_{10}\mathcal{M} \)) as a function of the spin precession parameter (\(\sqrt{(s_x)^2 + (s_y)^2}\)) for varying numbers of signals (2 to 5), demonstrating the robustness of our model under different levels of spin precession. (b) The mismatch (\( \log_{10}\mathcal{M} \)) as a function of the initial orbital eccentricity (\(e_0\)), highlighting the model's ability to generalize waveforms of varying levels of orbital eccentricity. Both subfigures use box plots to summarize the variability in mismatch values. The consistently low mismatch across diverse physical conditions illustrates the strong generalization capability of our model in separating signals with varying parameters.
	}
	\label{fig:gen_stat}
\end{figure*}

\subsection{Generalization Ability}
% SEOBNRE Test
Generalization is a critical property for models in gravitational wave data analysis, enabling them to perform well on unseen scenarios and diverse physical conditions. To explore the generalization ability of our model, we provide an example involving the separation of five overlapping signals burried in noise in \Cref{fig:generalize}. These signals include 3 \ac{BBH} waveforms with spin precession, orbital eccentricity, and higher-order modes, as well as \ac{NSBH} and \ac{BNS} waveforms. The waveform with spin precession and orbital eccentricity is generated by \texttt{SEOBNRE} \cite{liu_effectiv_2024,cao_waveform_2017}. The results demonstrate that our model can effectively separate and reconstruct diverse types of signals, including \ac{BBH}, \ac{NSBH}, and \ac{BNS} waveforms, even under challenging conditions. This highlights the model's robust generalization performance across a wide range of astrophysical scenarios and signal characteristics.

Our model demonstrates robust generalization to signals with varying precession parameters, even when applied to waveforms outside the training data. \Cref{fig:p_mismatch} shows the mismatch (\( \log_{10}\mathcal{M} \)) as a function of the precession parameter (\(\sqrt{(s^x)^2 + (s^y)^2}\)) for different numbers of sources. Here, we set $s^z = \sqrt{1-(s_x)^2 - (s_y)^2}$. The mismatch values remain consistently low, with only a slight performance drop as the precession parameter increases, indicating the model's strong ability to generalize to complex precessional dynamics while maintaining high signal fidelity.

The model also generalizes to signals under various orbital eccentricities, with \(e_0\) sampled from the range of \([0, 0.5]\). \Cref{fig:e0_mismatch} presents the mismatch (\( \log_{10}\mathcal{M} \)) as a function of the initial orbital eccentricity (\(e_0\)). The results show that the mismatch remains low across all eccentricity values tested, including scenarios where \(e_0\) deviates significantly from circular orbits. These findings indicate that the model can accurately reconstruct signals with complex orbital dynamics, further underscoring its versatility in handling diverse astrophysical phenomena.

In addition, although our model was trained on datasets containing between two and five signals, we tested its performance on data with only one signal present, effectively reducing the task to denoising rather than multi-source separation. As showcased in \Cref{fig:denoise}, the UnMixFormer achieves an overlap of 0.9969 in this single-signal scenario, highlighting its robust ability to extract and denoise even when confronted with only one source. Furthermore, because 3rd-generation detectors can record waveforms longer than one second, the data fed into our model may primarily contain inspiral-only segments of CBC signals. To accommodate this, we incorporated inspiral-only waveforms into our training dataset and retrained the model. As illustrated in \Cref{fig:inspiral}, our model continues to perform robustly, delivering accurate separation results even under these conditions.

\section{\label{sec:discussion} Conclusion and Discussion}

The detection and separation of overlapping \ac{CBC} \ac{GW} signals are pivotal challenges in \ac{GW} astronomy.
In this work, we introduced the UnMixFormer, an innovative neural network tailored for counting and separating overlapping sources.
By leveraging a dual-path architecture with attention-based blocks and \ac{FAN}, our model achieves high performance in handling complex overlapping signals. Evaluations on synthetic datasets, simulated with \ac{CE} noise, demonstrate robust number-of-signal counting and generalization ability, as well as high-fidelity waveform separation at varying \ac{SNR} levels. These results underscore the efficacy of our model and highlight promising avenues for future research.

One natural extension of this work lies in adapting the model for multi-detector networks. Incorporating data from multiple detectors such as \ac{CE}, \ac{ET} and \ac{LIGO} could improve source localization and separation by utilizing spatial diversity. This extension would require adapting the architecture to simultaneously process multi-channel data and integrate spatial information from antenna pattern functions across detectors. A multi-detector framework would significantly enhance the robustness and scalability of our model for real-world \ac{GW} observations.

Another promising direction is the application of our model to space-based \ac{GW} observatories, such as \ac{LISA}. For these missions, the detection and separation of \ac{MBHB} signals present unique challenges. Beyond \ac{MBHB}, the model could also be extended to address the separation of \ac{GB} foreground signals, which requires precise frequency-domain analysis due to their dense spectral distribution. Such adaptations would demonstrate the versatility of our model in handling the diverse signal scenarios expected from space-based observatories.

In conclusion, the UnMixFormer represents an advancement in \ac{GW} data analysis, providing a robust framework for tackling overlapping \ac{CBC} signals. While the current implementation demonstrates strong performance, extending it to multi-detector networks and space-based applications will ensure its continued relevance and impact in the evolving field of GW astronomy.

\begin{acknowledgments}
	This research was partially supported by the National Key Research and Development Program of China (Grant No. 2021YFC2201901 and 2021YFC2203001) and the National Natural Science Foundation of China (NSFC, Grant No. 11920101003 and 12021003). Additional support was provided by the Peng Cheng Laboratory and the Peng Cheng Laboratory Cloud-Brain. P. Xu was supported by the ``International Partnership Program of the Chinese Academy of Sciences'' (Grant No. 025GJHZ2023106GC). Z. Cao was supported by the ``Fundamental Research Funds for the Central Universities''.

\end{acknowledgments}

\bibliographystyle{apsrev4-2}
\bibliography{ref}

%apsrev4-2.bst 2015-08-30 from 4.21a (PWD, AO, DPC/HNN) hacked
\begin{thebibliography}{66}%
\makeatletter
\providecommand \@ifxundefined [1]{%
 \@ifx{#1\undefined}
}%
\providecommand \@ifnum [1]{%
 \ifnum #1\expandafter \@firstoftwo
 \else \expandafter \@secondoftwo
 \fi
}%
\providecommand \@ifx [1]{%
 \ifx #1\expandafter \@firstoftwo
 \else \expandafter \@secondoftwo
 \fi
}%
\providecommand \natexlab [1]{#1}%
\providecommand \enquote  [1]{``#1''}%
\providecommand \bibnamefont  [1]{#1}%
\providecommand \bibfnamefont [1]{#1}%
\providecommand \citenamefont [1]{#1}%
\providecommand \href@noop [0]{\@secondoftwo}%
\providecommand \href [0]{\begingroup \@sanitize@url \@href}%
\providecommand \@href[1]{\@@startlink{#1}\@@href}%
\providecommand \@@href[1]{\endgroup#1\@@endlink}%
\providecommand \@sanitize@url [0]{\catcode `\\12\catcode `\$12\catcode `\&12\catcode `\#12\catcode `\^12\catcode `\_12\catcode `\%12\relax}%
\providecommand \@@startlink[1]{}%
\providecommand \@@endlink[0]{}%
\providecommand \url  [0]{\begingroup\@sanitize@url \@url }%
\providecommand \@url [1]{\endgroup\@href {#1}{\urlprefix }}%
\providecommand \urlprefix  [0]{URL }%
\providecommand \Eprint [0]{\href }%
\providecommand \doibase [0]{http://dx.doi.org/}%
\providecommand \selectlanguage [0]{\@gobble}%
\providecommand \bibinfo  [0]{\@secondoftwo}%
\providecommand \bibfield  [0]{\@secondoftwo}%
\providecommand \translation [1]{[#1]}%
\providecommand \BibitemOpen [0]{}%
\providecommand \bibitemStop [0]{}%
\providecommand \bibitemNoStop [0]{.\EOS\space}%
\providecommand \EOS [0]{\spacefactor3000\relax}%
\providecommand \BibitemShut  [1]{\csname bibitem#1\endcsname}%
\let\auto@bib@innerbib\@empty
%</preamble>
\bibitem [{\citenamefont {Abbott}\ \emph {et~al.}(2016)\citenamefont {Abbott}, \citenamefont {Abbott}, \citenamefont {Abbott}, \citenamefont {Abernathy}, \citenamefont {Acernese}, \citenamefont {Ackley}, \citenamefont {Adams}, \citenamefont {Adams}, \citenamefont {Addesso}, \citenamefont {Adhikari}, \citenamefont {Adya}, \citenamefont {Affeldt}, \citenamefont {Agathos}, \citenamefont {Agatsuma}, \citenamefont {Aggarwal}, \citenamefont {Aguiar}, \citenamefont {Aiello}, \citenamefont {Ain}, \citenamefont {Ajith}, \citenamefont {Allen}, \citenamefont {Allocca}, \citenamefont {Altin}, \citenamefont {Anderson}, \citenamefont {Anderson}, \citenamefont {Arai}, \citenamefont {Arain}, \citenamefont {Araya}, \citenamefont {Arceneaux}, \citenamefont {Areeda}, \citenamefont {Arnaud}, \citenamefont {Arun}, \citenamefont {Ascenzi}, \citenamefont {Ashton}, \citenamefont {Ast}, \citenamefont {Aston}, \citenamefont {Astone}, \citenamefont {Aufmuth}, \citenamefont {Aulbert}, \citenamefont {Babak}, \citenamefont {Bacon},
  \citenamefont {Bader}, \citenamefont {Baker}, \citenamefont {Baldaccini}, \citenamefont {Ballardin}, \citenamefont {Ballmer}, \citenamefont {Barayoga}, \citenamefont {Barclay}, \citenamefont {Barish}, \citenamefont {Barker}, \citenamefont {Barone}, \citenamefont {Barr}, \citenamefont {Barsotti}, \citenamefont {Barsuglia}, \citenamefont {Barta}, \citenamefont {Bartlett}, \citenamefont {Barton}, \citenamefont {Bartos}, \citenamefont {Bassiri}, \citenamefont {Basti}, \citenamefont {Batch}, \citenamefont {Baune}, \citenamefont {Bavigadda}, \citenamefont {Bazzan}, \citenamefont {Behnke}, \citenamefont {Bejger}, \citenamefont {Belczynski}, \citenamefont {Bell}, \citenamefont {Bell}, \citenamefont {Berger}, \citenamefont {Bergman}, \citenamefont {Bergmann}, \citenamefont {Berry}, \citenamefont {Bersanetti}, \citenamefont {Bertolini}, \citenamefont {Betzwieser}, \citenamefont {Bhagwat}, \citenamefont {Bhandare}, \citenamefont {Bilenko}, \citenamefont {Billingsley}, \citenamefont {Birch}, \citenamefont {Birney},
  \citenamefont {Birnholtz}, \citenamefont {Biscans}, \citenamefont {Bisht}, \citenamefont {Bitossi}, \citenamefont {Biwer}, \citenamefont {Bizouard}, \citenamefont {Blackburn}, \citenamefont {Blair}, \citenamefont {Blair}, \citenamefont {Blair}, \citenamefont {Bloemen}, \citenamefont {Bock}, \citenamefont {Bodiya}, \citenamefont {Boer}, \citenamefont {Bogaert}, \citenamefont {Bogan}, \citenamefont {Bohe}, \citenamefont {Bojtos}, \citenamefont {Bond}, \citenamefont {Bondu}, \citenamefont {Bonnand}, \citenamefont {Boom}, \citenamefont {Bork}, \citenamefont {Boschi}, \citenamefont {Bose}, \citenamefont {Bouffanais}, \citenamefont {Bozzi}, \citenamefont {Bradaschia}, \citenamefont {Brady}, \citenamefont {Braginsky}, \citenamefont {Branchesi}, \citenamefont {Brau}, \citenamefont {Briant}, \citenamefont {Brillet}, \citenamefont {Brinkmann}, \citenamefont {Brisson}, \citenamefont {Brockill}, \citenamefont {Brooks}, \citenamefont {Brown}, \citenamefont {Brown}, \citenamefont {Brown}, \citenamefont {Buchanan},
  \citenamefont {Buikema}, \citenamefont {Bulik}, \citenamefont {Bulten}, \citenamefont {Buonanno}, \citenamefont {Buskulic}, \citenamefont {Buy}, \citenamefont {Byer}, \citenamefont {Cabero}, \citenamefont {Cadonati}, \citenamefont {Cagnoli}, \citenamefont {Cahillane}, \citenamefont {Bustillo}, \citenamefont {Callister}, \citenamefont {Calloni}, \citenamefont {Camp}, \citenamefont {Cannon}, \citenamefont {Cao}, \citenamefont {Capano}, \citenamefont {Capocasa}, \citenamefont {Carbognani}, \citenamefont {Caride}, \citenamefont {Diaz}, \citenamefont {Casentini}, \citenamefont {Caudill}, \citenamefont {Cavagli{\`a}}, \citenamefont {Cavalier}, \citenamefont {Cavalieri}, \citenamefont {Cella}, \citenamefont {Cepeda}, \citenamefont {Baiardi}, \citenamefont {Cerretani}, \citenamefont {Cesarini}, \citenamefont {Chakraborty}, \citenamefont {Chalermsongsak}, \citenamefont {Chamberlin}, \citenamefont {Chan}, \citenamefont {Chao}, \citenamefont {Charlton}, \citenamefont {{Chassande-Mottin}}, \citenamefont {Chen},
  \citenamefont {Chen}, \citenamefont {Cheng}, \citenamefont {Chincarini}, \citenamefont {Chiummo}, \citenamefont {Cho}, \citenamefont {Cho}, \citenamefont {Chow}, \citenamefont {Christensen}, \citenamefont {Chu}, \citenamefont {Chua}, \citenamefont {Chung}, \citenamefont {Ciani}, \citenamefont {Clara}, \citenamefont {Clark}, \citenamefont {Cleva}, \citenamefont {Coccia}, \citenamefont {Cohadon}, \citenamefont {Colla}, \citenamefont {Collette}, \citenamefont {Cominsky}, \citenamefont {Constancio}, \citenamefont {Conte}, \citenamefont {Conti}, \citenamefont {Cook}, \citenamefont {Corbitt}, \citenamefont {Cornish}, \citenamefont {Corsi}, \citenamefont {Cortese}, \citenamefont {Costa}, \citenamefont {Coughlin}, \citenamefont {Coughlin}, \citenamefont {Coulon}, \citenamefont {Countryman}, \citenamefont {Couvares}, \citenamefont {Cowan}, \citenamefont {Coward}, \citenamefont {Cowart}, \citenamefont {Coyne}, \citenamefont {Coyne}, \citenamefont {Craig}, \citenamefont {Creighton}, \citenamefont {Creighton},
  \citenamefont {Cripe}, \citenamefont {Crowder}, \citenamefont {Cruise}, \citenamefont {Cumming}, \citenamefont {Cunningham}, \citenamefont {Cuoco}, \citenamefont {Canton}, \citenamefont {Danilishin}, \citenamefont {D'Antonio}, \citenamefont {Danzmann}, \citenamefont {Darman}, \citenamefont {Da~Silva~Costa}, \citenamefont {Dattilo}, \citenamefont {Dave}, \citenamefont {Daveloza}, \citenamefont {Davier}, \citenamefont {Davies}, \citenamefont {Daw}, \citenamefont {Day}, \citenamefont {De}, \citenamefont {DeBra}, \citenamefont {Debreczeni}, \citenamefont {Degallaix}, \citenamefont {De~Laurentis}, \citenamefont {Del{\'e}glise}, \citenamefont {Del~Pozzo}, \citenamefont {Denker}, \citenamefont {Dent}, \citenamefont {Dereli}, \citenamefont {Dergachev}, \citenamefont {DeRosa}, \citenamefont {De~Rosa}, \citenamefont {DeSalvo}, \citenamefont {Dhurandhar}, \citenamefont {D{\'i}az}, \citenamefont {Di~Fiore}, \citenamefont {Di~Giovanni}, \citenamefont {Di~Lieto}, \citenamefont {Di~Pace}, \citenamefont {Di~Palma},
  \citenamefont {Di~Virgilio}, \citenamefont {Dojcinoski}, \citenamefont {Dolique}, \citenamefont {Donovan}, \citenamefont {Dooley}, \citenamefont {Doravari}, \citenamefont {Douglas}, \citenamefont {Downes}, \citenamefont {Drago}, \citenamefont {Drever}, \citenamefont {Driggers}, \citenamefont {Du}, \citenamefont {Ducrot}, \citenamefont {Dwyer}, \citenamefont {Edo}, \citenamefont {Edwards}, \citenamefont {Effler}, \citenamefont {Eggenstein}, \citenamefont {Ehrens}, \citenamefont {Eichholz}, \citenamefont {Eikenberry}, \citenamefont {Engels}, \citenamefont {Essick}, \citenamefont {Etzel}, \citenamefont {Evans}, \citenamefont {Evans}, \citenamefont {Everett}, \citenamefont {Factourovich}, \citenamefont {Fafone}, \citenamefont {Fair}, \citenamefont {Fairhurst}, \citenamefont {Fan}, \citenamefont {Fang}, \citenamefont {Farinon}, \citenamefont {Farr}, \citenamefont {Farr}, \citenamefont {Favata}, \citenamefont {Fays}, \citenamefont {Fehrmann}, \citenamefont {Fejer}, \citenamefont {Feldbaum}, \citenamefont
  {Ferrante}, \citenamefont {Ferreira}, \citenamefont {Ferrini}, \citenamefont {Fidecaro}, \citenamefont {Finn}, \citenamefont {Fiori}, \citenamefont {Fiorucci}, \citenamefont {Fisher}, \citenamefont {Flaminio}, \citenamefont {Fletcher}, \citenamefont {Fong}, \citenamefont {Fournier}, \citenamefont {Franco}, \citenamefont {Frasca}, \citenamefont {Frasconi}, \citenamefont {Frede}, \citenamefont {Frei}, \citenamefont {Freise}, \citenamefont {Frey}, \citenamefont {Frey}, \citenamefont {Fricke}, \citenamefont {Fritschel}, \citenamefont {Frolov}, \citenamefont {Fulda}, \citenamefont {Fyffe}, \citenamefont {Gabbard}, \citenamefont {Gair}, \citenamefont {Gammaitoni}, \citenamefont {Gaonkar}, \citenamefont {Garufi}, \citenamefont {Gatto}, \citenamefont {Gaur}, \citenamefont {Gehrels}, \citenamefont {Gemme}, \citenamefont {Gendre}, \citenamefont {Genin}, \citenamefont {Gennai}, \citenamefont {George}, \citenamefont {Gergely}, \citenamefont {Germain}, \citenamefont {Ghosh}, \citenamefont {Ghosh}, \citenamefont {Ghosh},
  \citenamefont {Giaime}, \citenamefont {Giardina}, \citenamefont {Giazotto}, \citenamefont {Gill}, \citenamefont {Glaefke}, \citenamefont {Gleason}, \citenamefont {Goetz}, \citenamefont {Goetz}, \citenamefont {Gondan}, \citenamefont {Gonz{\'a}lez}, \citenamefont {Castro}, \citenamefont {Gopakumar}, \citenamefont {Gordon}, \citenamefont {Gorodetsky}, \citenamefont {Gossan}, \citenamefont {Gosselin}, \citenamefont {Gouaty}, \citenamefont {Graef}, \citenamefont {Graff}, \citenamefont {Granata}, \citenamefont {Grant}, \citenamefont {Gras}, \citenamefont {Gray}, \citenamefont {Greco}, \citenamefont {Green}, \citenamefont {Greenhalgh}, \citenamefont {Groot}, \citenamefont {Grote}, \citenamefont {Grunewald}, \citenamefont {Guidi}, \citenamefont {Guo}, \citenamefont {Gupta}, \citenamefont {Gupta}, \citenamefont {Gushwa}, \citenamefont {Gustafson}, \citenamefont {Gustafson}, \citenamefont {Hacker}, \citenamefont {Hall}, \citenamefont {Hall}, \citenamefont {Hammond}, \citenamefont {Haney}, \citenamefont {Hanke},
  \citenamefont {Hanks}, \citenamefont {Hanna}, \citenamefont {Hannam}, \citenamefont {Hanson}, \citenamefont {Hardwick}, \citenamefont {Harms}, \citenamefont {Harry}, \citenamefont {Harry}, \citenamefont {Hart}, \citenamefont {Hartman}, \citenamefont {Haster}, \citenamefont {Haughian}, \citenamefont {Healy}, \citenamefont {Heefner}, \citenamefont {Heidmann}, \citenamefont {Heintze}, \citenamefont {Heinzel}, \citenamefont {Heitmann}, \citenamefont {Hello}, \citenamefont {Hemming}, \citenamefont {Hendry}, \citenamefont {Heng}, \citenamefont {Hennig}, \citenamefont {Heptonstall}, \citenamefont {Heurs}, \citenamefont {Hild}, \citenamefont {Hoak}, \citenamefont {Hodge}, \citenamefont {Hofman}, \citenamefont {Hollitt}, \citenamefont {Holt}, \citenamefont {Holz}, \citenamefont {Hopkins}, \citenamefont {Hosken}, \citenamefont {Hough}, \citenamefont {Houston}, \citenamefont {Howell}, \citenamefont {Hu}, \citenamefont {Huang}, \citenamefont {Huerta}, \citenamefont {Huet}, \citenamefont {Hughey}, \citenamefont {Husa},
  \citenamefont {Huttner}, \citenamefont {{Huynh-Dinh}}, \citenamefont {Idrisy}, \citenamefont {Indik}, \citenamefont {Ingram}, \citenamefont {Inta}, \citenamefont {Isa}, \citenamefont {Isac}, \citenamefont {Isi}, \citenamefont {Islas}, \citenamefont {Isogai}, \citenamefont {Iyer}, \citenamefont {Izumi}, \citenamefont {Jacobson}, \citenamefont {Jacqmin}, \citenamefont {Jang}, \citenamefont {Jani}, \citenamefont {Jaranowski}, \citenamefont {Jawahar}, \citenamefont {{Jim{\'e}nez-Forteza}}, \citenamefont {Johnson}, \citenamefont {{Johnson-McDaniel}}, \citenamefont {Jones}, \citenamefont {Jones}, \citenamefont {Jonker}, \citenamefont {Ju}, \citenamefont {Haris}, \citenamefont {Kalaghatgi}, \citenamefont {Kalogera}, \citenamefont {Kandhasamy}, \citenamefont {Kang}, \citenamefont {Kanner}, \citenamefont {Karki}, \citenamefont {Kasprzack}, \citenamefont {Katsavounidis}, \citenamefont {Katzman}, \citenamefont {Kaufer}, \citenamefont {Kaur}, \citenamefont {Kawabe}, \citenamefont {Kawazoe}, \citenamefont
  {K{\'e}f{\'e}lian}, \citenamefont {Kehl}, \citenamefont {Keitel}, \citenamefont {Kelley}, \citenamefont {Kells}, \citenamefont {Kennedy}, \citenamefont {Keppel}, \citenamefont {Key}, \citenamefont {Khalaidovski}, \citenamefont {Khalili}, \citenamefont {Khan}, \citenamefont {Khan}, \citenamefont {Khan}, \citenamefont {Khazanov}, \citenamefont {Kijbunchoo}, \citenamefont {Kim}, \citenamefont {Kim}, \citenamefont {Kim}, \citenamefont {Kim}, \citenamefont {Kim}, \citenamefont {Kim}, \citenamefont {King}, \citenamefont {King}, \citenamefont {Kinzel}, \citenamefont {Kissel}, \citenamefont {Kleybolte}, \citenamefont {Klimenko}, \citenamefont {Koehlenbeck}, \citenamefont {Kokeyama}, \citenamefont {Koley}, \citenamefont {Kondrashov}, \citenamefont {Kontos}, \citenamefont {Koranda}, \citenamefont {Korobko}, \citenamefont {Korth}, \citenamefont {Kowalska}, \citenamefont {Kozak}, \citenamefont {Kringel}, \citenamefont {Krishnan}, \citenamefont {Kr{\'o}lak}, \citenamefont {Krueger}, \citenamefont {Kuehn}, \citenamefont
  {Kumar}, \citenamefont {Kumar}, \citenamefont {Kuo}, \citenamefont {Kutynia}, \citenamefont {Kwee}, \citenamefont {Lackey}, \citenamefont {Landry}, \citenamefont {Lange}, \citenamefont {Lantz}, \citenamefont {Lasky}, \citenamefont {Lazzarini}, \citenamefont {Lazzaro}, \citenamefont {Leaci}, \citenamefont {Leavey}, \citenamefont {Lebigot}, \citenamefont {Lee}, \citenamefont {Lee}, \citenamefont {Lee}, \citenamefont {Lee}, \citenamefont {Lenon}, \citenamefont {Leonardi}, \citenamefont {Leong}, \citenamefont {Leroy}, \citenamefont {Letendre}, \citenamefont {Levin}, \citenamefont {Levine}, \citenamefont {Li}, \citenamefont {Libson}, \citenamefont {Littenberg}, \citenamefont {Lockerbie}, \citenamefont {Logue}, \citenamefont {Lombardi}, \citenamefont {London}, \citenamefont {Lord}, \citenamefont {Lorenzini}, \citenamefont {Loriette}, \citenamefont {Lormand}, \citenamefont {Losurdo}, \citenamefont {Lough}, \citenamefont {Lousto}, \citenamefont {Lovelace}, \citenamefont {L{\"u}ck}, \citenamefont {Lundgren},
  \citenamefont {Luo}, \citenamefont {Lynch}, \citenamefont {Ma}, \citenamefont {MacDonald}, \citenamefont {Machenschalk}, \citenamefont {MacInnis}, \citenamefont {Macleod}, \citenamefont {{Maga{\~n}a-Sandoval}}, \citenamefont {Magee}, \citenamefont {Mageswaran}, \citenamefont {Majorana}, \citenamefont {Maksimovic}, \citenamefont {Malvezzi}, \citenamefont {Man}, \citenamefont {Mandel}, \citenamefont {Mandic}, \citenamefont {Mangano}, \citenamefont {Mansell}, \citenamefont {Manske}, \citenamefont {Mantovani}, \citenamefont {Marchesoni}, \citenamefont {Marion}, \citenamefont {M{\'a}rka}, \citenamefont {M{\'a}rka}, \citenamefont {Markosyan}, \citenamefont {Maros}, \citenamefont {Martelli}, \citenamefont {Martellini}, \citenamefont {Martin}, \citenamefont {Martin}, \citenamefont {Martynov}, \citenamefont {Marx}, \citenamefont {Mason}, \citenamefont {Masserot}, \citenamefont {Massinger}, \citenamefont {{Masso-Reid}}, \citenamefont {Matichard}, \citenamefont {Matone}, \citenamefont {Mavalvala}, \citenamefont
  {Mazumder}, \citenamefont {Mazzolo}, \citenamefont {McCarthy}, \citenamefont {McClelland}, \citenamefont {McCormick}, \citenamefont {McGuire}, \citenamefont {McIntyre}, \citenamefont {McIver}, \citenamefont {McManus}, \citenamefont {McWilliams}, \citenamefont {Meacher}, \citenamefont {Meadors}, \citenamefont {Meidam}, \citenamefont {Melatos}, \citenamefont {Mendell}, \citenamefont {{Mendoza-Gandara}}, \citenamefont {Mercer}, \citenamefont {Merilh}, \citenamefont {Merzougui}, \citenamefont {Meshkov}, \citenamefont {Messenger}, \citenamefont {Messick}, \citenamefont {Meyers}, \citenamefont {Mezzani}, \citenamefont {Miao}, \citenamefont {Michel}, \citenamefont {Middleton}, \citenamefont {Mikhailov}, \citenamefont {Milano}, \citenamefont {Miller}, \citenamefont {Millhouse}, \citenamefont {Minenkov}, \citenamefont {Ming}, \citenamefont {Mirshekari}, \citenamefont {Mishra}, \citenamefont {Mitra}, \citenamefont {Mitrofanov}, \citenamefont {Mitselmakher}, \citenamefont {Mittleman}, \citenamefont {Moggi},
  \citenamefont {Mohan}, \citenamefont {Mohapatra}, \citenamefont {Montani}, \citenamefont {Moore}, \citenamefont {Moore}, \citenamefont {Moraru}, \citenamefont {Moreno}, \citenamefont {Morriss}, \citenamefont {Mossavi}, \citenamefont {Mours}, \citenamefont {{Mow-Lowry}}, \citenamefont {Mueller}, \citenamefont {Mueller}, \citenamefont {Muir}, \citenamefont {Mukherjee}, \citenamefont {Mukherjee}, \citenamefont {Mukherjee}, \citenamefont {Mukund}, \citenamefont {Mullavey}, \citenamefont {Munch}, \citenamefont {Murphy}, \citenamefont {Murray}, \citenamefont {Mytidis}, \citenamefont {Nardecchia}, \citenamefont {Naticchioni}, \citenamefont {Nayak}, \citenamefont {Necula}, \citenamefont {Nedkova}, \citenamefont {Nelemans}, \citenamefont {Neri}, \citenamefont {Neunzert}, \citenamefont {Newton}, \citenamefont {Nguyen}, \citenamefont {Nielsen}, \citenamefont {Nissanke}, \citenamefont {Nitz}, \citenamefont {Nocera}, \citenamefont {Nolting}, \citenamefont {Normandin}, \citenamefont {Nuttall}, \citenamefont {Oberling},
  \citenamefont {Ochsner}, \citenamefont {O'Dell}, \citenamefont {Oelker}, \citenamefont {Ogin}, \citenamefont {Oh}, \citenamefont {Oh}, \citenamefont {Ohme}, \citenamefont {Oliver}, \citenamefont {Oppermann}, \citenamefont {Oram}, \citenamefont {O'Reilly}, \citenamefont {O'Shaughnessy}, \citenamefont {Ott}, \citenamefont {Ottaway}, \citenamefont {Ottens}, \citenamefont {Overmier}, \citenamefont {Owen}, \citenamefont {Pai}, \citenamefont {Pai}, \citenamefont {Palamos}, \citenamefont {Palashov}, \citenamefont {Palomba}, \citenamefont {{Pal-Singh}}, \citenamefont {Pan}, \citenamefont {Pan}, \citenamefont {Pankow}, \citenamefont {Pannarale}, \citenamefont {Pant}, \citenamefont {Paoletti}, \citenamefont {Paoli}, \citenamefont {Papa}, \citenamefont {Paris}, \citenamefont {Parker}, \citenamefont {Pascucci}, \citenamefont {Pasqualetti}, \citenamefont {Passaquieti}, \citenamefont {Passuello}, \citenamefont {Patricelli}, \citenamefont {Patrick}, \citenamefont {Pearlstone}, \citenamefont {Pedraza}, \citenamefont
  {Pedurand}, \citenamefont {Pekowsky}, \citenamefont {Pele}, \citenamefont {Penn}, \citenamefont {Perreca}, \citenamefont {Pfeiffer}, \citenamefont {Phelps}, \citenamefont {Piccinni}, \citenamefont {Pichot}, \citenamefont {Pickenpack}, \citenamefont {Piergiovanni}, \citenamefont {Pierro}, \citenamefont {Pillant}, \citenamefont {Pinard}, \citenamefont {Pinto}, \citenamefont {Pitkin}, \citenamefont {Poeld}, \citenamefont {Poggiani}, \citenamefont {Popolizio}, \citenamefont {Post}, \citenamefont {Powell}, \citenamefont {Prasad}, \citenamefont {Predoi}, \citenamefont {Premachandra}, \citenamefont {Prestegard}, \citenamefont {Price}, \citenamefont {Prijatelj}, \citenamefont {Principe}, \citenamefont {Privitera}, \citenamefont {Prix}, \citenamefont {Prodi}, \citenamefont {Prokhorov}, \citenamefont {Puncken}, \citenamefont {Punturo}, \citenamefont {Puppo}, \citenamefont {P{\"u}rrer}, \citenamefont {Qi}, \citenamefont {Qin}, \citenamefont {Quetschke}, \citenamefont {Quintero}, \citenamefont {{Quitzow-James}},
  \citenamefont {Raab}, \citenamefont {Rabeling}, \citenamefont {Radkins}, \citenamefont {Raffai}, \citenamefont {Raja}, \citenamefont {Rakhmanov}, \citenamefont {Ramet}, \citenamefont {Rapagnani}, \citenamefont {Raymond}, \citenamefont {Razzano}, \citenamefont {Re}, \citenamefont {Read}, \citenamefont {Reed}, \citenamefont {Regimbau}, \citenamefont {Rei}, \citenamefont {Reid}, \citenamefont {Reitze}, \citenamefont {Rew}, \citenamefont {Reyes}, \citenamefont {Ricci}, \citenamefont {Riles}, \citenamefont {Robertson}, \citenamefont {Robie}, \citenamefont {Robinet}, \citenamefont {Rocchi}, \citenamefont {Rolland}, \citenamefont {Rollins}, \citenamefont {Roma}, \citenamefont {Romano}, \citenamefont {Romano}, \citenamefont {Romanov}, \citenamefont {Romie}, \citenamefont {Rosi{\'n}ska}, \citenamefont {Rowan}, \citenamefont {R{\"u}diger}, \citenamefont {Ruggi}, \citenamefont {Ryan}, \citenamefont {Sachdev}, \citenamefont {Sadecki}, \citenamefont {Sadeghian}, \citenamefont {Salconi}, \citenamefont {Saleem},
  \citenamefont {Salemi}, \citenamefont {Samajdar}, \citenamefont {Sammut}, \citenamefont {Sampson}, \citenamefont {Sanchez}, \citenamefont {Sandberg}, \citenamefont {Sandeen}, \citenamefont {Sanders}, \citenamefont {Sanders}, \citenamefont {Sassolas}, \citenamefont {Sathyaprakash}, \citenamefont {Saulson}, \citenamefont {Sauter}, \citenamefont {Savage}, \citenamefont {Sawadsky}, \citenamefont {Schale}, \citenamefont {Schilling}, \citenamefont {Schmidt}, \citenamefont {Schmidt}, \citenamefont {Schnabel}, \citenamefont {Schofield}, \citenamefont {Sch{\"o}nbeck}, \citenamefont {Schreiber}, \citenamefont {Schuette}, \citenamefont {Schutz}, \citenamefont {Scott}, \citenamefont {Scott}, \citenamefont {Sellers}, \citenamefont {Sengupta}, \citenamefont {Sentenac}, \citenamefont {Sequino}, \citenamefont {Sergeev}, \citenamefont {Serna}, \citenamefont {Setyawati}, \citenamefont {Sevigny}, \citenamefont {Shaddock}, \citenamefont {Shaffer}, \citenamefont {Shah}, \citenamefont {Shahriar}, \citenamefont {Shaltev},
  \citenamefont {Shao}, \citenamefont {Shapiro}, \citenamefont {Shawhan}, \citenamefont {Sheperd}, \citenamefont {Shoemaker}, \citenamefont {Shoemaker}, \citenamefont {Siellez}, \citenamefont {Siemens}, \citenamefont {Sigg}, \citenamefont {Silva}, \citenamefont {Simakov}, \citenamefont {Singer}, \citenamefont {Singer}, \citenamefont {Singh}, \citenamefont {Singh}, \citenamefont {Singhal}, \citenamefont {Sintes}, \citenamefont {Slagmolen}, \citenamefont {Smith}, \citenamefont {Smith}, \citenamefont {Smith}, \citenamefont {Smith}, \citenamefont {Son}, \citenamefont {Sorazu}, \citenamefont {Sorrentino}, \citenamefont {Souradeep}, \citenamefont {Srivastava}, \citenamefont {Staley}, \citenamefont {Steinke}, \citenamefont {Steinlechner}, \citenamefont {Steinlechner}, \citenamefont {Steinmeyer}, \citenamefont {Stephens}, \citenamefont {Stevenson}, \citenamefont {Stone}, \citenamefont {Strain}, \citenamefont {Straniero}, \citenamefont {Stratta}, \citenamefont {Strauss}, \citenamefont {Strigin}, \citenamefont
  {Sturani}, \citenamefont {Stuver}, \citenamefont {Summerscales}, \citenamefont {Sun}, \citenamefont {Sutton}, \citenamefont {Swinkels}, \citenamefont {Szczepa{\'n}czyk}, \citenamefont {Tacca}, \citenamefont {Talukder}, \citenamefont {Tanner}, \citenamefont {T{\'a}pai}, \citenamefont {Tarabrin}, \citenamefont {Taracchini}, \citenamefont {Taylor}, \citenamefont {Theeg}, \citenamefont {Thirugnanasambandam}, \citenamefont {Thomas}, \citenamefont {Thomas}, \citenamefont {Thomas}, \citenamefont {Thorne}, \citenamefont {Thorne}, \citenamefont {Thrane}, \citenamefont {Tiwari}, \citenamefont {Tiwari}, \citenamefont {Tokmakov}, \citenamefont {Tomlinson}, \citenamefont {Tonelli}, \citenamefont {Torres}, \citenamefont {Torrie}, \citenamefont {T{\"o}yr{\"a}}, \citenamefont {Travasso}, \citenamefont {Traylor}, \citenamefont {Trifir{\`o}}, \citenamefont {Tringali}, \citenamefont {Trozzo}, \citenamefont {Tse}, \citenamefont {Turconi}, \citenamefont {Tuyenbayev}, \citenamefont {Ugolini}, \citenamefont {Unnikrishnan},
  \citenamefont {Urban}, \citenamefont {Usman}, \citenamefont {Vahlbruch}, \citenamefont {Vajente}, \citenamefont {Valdes}, \citenamefont {Vallisneri}, \citenamefont {Van~Bakel}, \citenamefont {Van~Beuzekom}, \citenamefont {Van Den~Brand}, \citenamefont {Van Den~Broeck}, \citenamefont {{Vander-Hyde}}, \citenamefont {Van Der~Schaaf}, \citenamefont {Van~Heijningen}, \citenamefont {Van~Veggel}, \citenamefont {Vardaro}, \citenamefont {Vass}, \citenamefont {Vas{\'u}th}, \citenamefont {Vaulin}, \citenamefont {Vecchio}, \citenamefont {Vedovato}, \citenamefont {Veitch}, \citenamefont {Veitch}, \citenamefont {Venkateswara}, \citenamefont {Verkindt}, \citenamefont {Vetrano}, \citenamefont {Vicer{\'e}}, \citenamefont {Vinciguerra}, \citenamefont {Vine}, \citenamefont {Vinet}, \citenamefont {Vitale}, \citenamefont {Vo}, \citenamefont {Vocca}, \citenamefont {Vorvick}, \citenamefont {Voss}, \citenamefont {Vousden}, \citenamefont {Vyatchanin}, \citenamefont {Wade}, \citenamefont {Wade}, \citenamefont {Wade}, \citenamefont
  {Waldman}, \citenamefont {Walker}, \citenamefont {Wallace}, \citenamefont {Walsh}, \citenamefont {Wang}, \citenamefont {Wang}, \citenamefont {Wang}, \citenamefont {Wang}, \citenamefont {Wang}, \citenamefont {Ward}, \citenamefont {Ward}, \citenamefont {Warner}, \citenamefont {Was}, \citenamefont {Weaver}, \citenamefont {Wei}, \citenamefont {Weinert}, \citenamefont {Weinstein}, \citenamefont {Weiss}, \citenamefont {Welborn}, \citenamefont {Wen}, \citenamefont {We{\ss}els}, \citenamefont {Westphal}, \citenamefont {Wette}, \citenamefont {Whelan}, \citenamefont {Whitcomb}, \citenamefont {White}, \citenamefont {Whiting}, \citenamefont {Wiesner}, \citenamefont {Wilkinson}, \citenamefont {Willems}, \citenamefont {Williams}, \citenamefont {Williams}, \citenamefont {Williamson}, \citenamefont {Willis}, \citenamefont {Willke}, \citenamefont {Wimmer}, \citenamefont {Winkelmann}, \citenamefont {Winkler}, \citenamefont {Wipf}, \citenamefont {Wiseman}, \citenamefont {Wittel}, \citenamefont {Woan}, \citenamefont {Worden},
  \citenamefont {Wright}, \citenamefont {Wu}, \citenamefont {Yablon}, \citenamefont {Yakushin}, \citenamefont {Yam}, \citenamefont {Yamamoto}, \citenamefont {Yancey}, \citenamefont {Yap}, \citenamefont {Yu}, \citenamefont {Yvert}, \citenamefont {Zadro{\.z}ny}, \citenamefont {Zangrando}, \citenamefont {Zanolin}, \citenamefont {Zendri}, \citenamefont {Zevin}, \citenamefont {Zhang}, \citenamefont {Zhang}, \citenamefont {Zhang}, \citenamefont {Zhang}, \citenamefont {Zhao}, \citenamefont {Zhou}, \citenamefont {Zhou}, \citenamefont {Zhu}, \citenamefont {Zucker}, \citenamefont {Zuraw}, \citenamefont {Zweizig},\ and\ \citenamefont {{LIGO Scientific Collaboration and Virgo Collaboration}}}]{abbott_observation_2016}%
  \BibitemOpen
  \bibfield  {author} {\bibinfo {author} {\bibfnamefont {B.~P.}\ \bibnamefont {Abbott}}, \bibinfo {author} {\bibfnamefont {R.}~\bibnamefont {Abbott}}, \bibinfo {author} {\bibfnamefont {T.~D.}\ \bibnamefont {Abbott}}, \bibinfo {author} {\bibfnamefont {M.~R.}\ \bibnamefont {Abernathy}}, \bibinfo {author} {\bibfnamefont {F.}~\bibnamefont {Acernese}}, \bibnamefont {et~al.},\ }\href {\doibase 10.1103/PhysRevLett.116.061102} {\bibfield  {journal} {\bibinfo  {journal} {Phys. Rev. Lett.}\ }\textbf {\bibinfo {volume} {116}},\ \bibinfo {pages} {061102} (\bibinfo {year} {2016})}\BibitemShut {NoStop}%
\bibitem [{\citenamefont {Cahillane}\ and\ \citenamefont {Mansell}(2022)}]{cahillane_review_2022}%
  \BibitemOpen
  \bibfield  {author} {\bibinfo {author} {\bibfnamefont {C.}~\bibnamefont {Cahillane}}\ \bibnamefont {and}\ \bibinfo {author} {\bibfnamefont {G.}~\bibnamefont {Mansell}},\ }\href {\doibase 10.3390/galaxies10010036} {\bibfield  {journal} {\bibinfo  {journal} {Galaxies}\ }\textbf {\bibinfo {volume} {10}},\ \bibinfo {pages} {36} (\bibinfo {year} {2022})}\BibitemShut {NoStop}%
\bibitem [{\citenamefont {Martynov}\ \emph {et~al.}(2016)\citenamefont {Martynov}, \citenamefont {Hall}, \citenamefont {Abbott}, \citenamefont {Abbott}, \citenamefont {Abbott}, \citenamefont {Adams}, \citenamefont {Adhikari}, \citenamefont {Anderson}, \citenamefont {Anderson}, \citenamefont {Arai}, \citenamefont {Arain}, \citenamefont {Aston}, \citenamefont {Austin}, \citenamefont {Ballmer}, \citenamefont {Barbet}, \citenamefont {Barker}, \citenamefont {Barr}, \citenamefont {Barsotti}, \citenamefont {Bartlett}, \citenamefont {Barton}, \citenamefont {Bartos}, \citenamefont {Batch}, \citenamefont {Bell}, \citenamefont {Belopolski}, \citenamefont {Bergman}, \citenamefont {Betzwieser}, \citenamefont {Billingsley}, \citenamefont {Birch}, \citenamefont {Biscans}, \citenamefont {Biwer}, \citenamefont {Black}, \citenamefont {Blair}, \citenamefont {Bogan}, \citenamefont {Bond}, \citenamefont {Bork}, \citenamefont {Bridges}, \citenamefont {Brooks}, \citenamefont {Brown}, \citenamefont {Carbone}, \citenamefont
  {Celerier}, \citenamefont {Ciani}, \citenamefont {Clara}, \citenamefont {Cook}, \citenamefont {Countryman}, \citenamefont {Cowart}, \citenamefont {Coyne}, \citenamefont {Cumming}, \citenamefont {Cunningham}, \citenamefont {Damjanic}, \citenamefont {Dannenberg}, \citenamefont {Danzmann}, \citenamefont {Costa}, \citenamefont {Daw}, \citenamefont {DeBra}, \citenamefont {DeRosa}, \citenamefont {DeSalvo}, \citenamefont {Dooley}, \citenamefont {Doravari}, \citenamefont {Driggers}, \citenamefont {Dwyer}, \citenamefont {Effler}, \citenamefont {Etzel}, \citenamefont {Evans}, \citenamefont {Evans}, \citenamefont {Factourovich}, \citenamefont {Fair}, \citenamefont {Feldbaum}, \citenamefont {Fisher}, \citenamefont {Foley}, \citenamefont {Frede}, \citenamefont {Freise}, \citenamefont {Fritschel}, \citenamefont {Frolov}, \citenamefont {Fulda}, \citenamefont {Fyffe}, \citenamefont {Galdi}, \citenamefont {Giaime}, \citenamefont {Giardina}, \citenamefont {Gleason}, \citenamefont {Goetz}, \citenamefont {Gras}, \citenamefont
  {Gray}, \citenamefont {Greenhalgh}, \citenamefont {Grote}, \citenamefont {Guido}, \citenamefont {Gushwa}, \citenamefont {Gustafson}, \citenamefont {Gustafson}, \citenamefont {Hammond}, \citenamefont {Hanks}, \citenamefont {Hanson}, \citenamefont {Hardwick}, \citenamefont {Harry}, \citenamefont {Haughian}, \citenamefont {Heefner}, \citenamefont {Heintze}, \citenamefont {Heptonstall}, \citenamefont {Hoak}, \citenamefont {Hough}, \citenamefont {Ivanov}, \citenamefont {Izumi}, \citenamefont {Jacobson}, \citenamefont {James}, \citenamefont {Jones}, \citenamefont {Kandhasamy}, \citenamefont {Karki}, \citenamefont {Kasprzack}, \citenamefont {Kaufer}, \citenamefont {Kawabe}, \citenamefont {Kells}, \citenamefont {Kijbunchoo}, \citenamefont {King}, \citenamefont {King}, \citenamefont {Kinzel}, \citenamefont {Kissel}, \citenamefont {Kokeyama}, \citenamefont {Korth}, \citenamefont {Kuehn}, \citenamefont {Kwee}, \citenamefont {Landry}, \citenamefont {Lantz}, \citenamefont {Roux}, \citenamefont {Levine}, \citenamefont
  {Lewis}, \citenamefont {Lhuillier}, \citenamefont {Lockerbie}, \citenamefont {Lormand}, \citenamefont {Lubinski}, \citenamefont {Lundgren}, \citenamefont {MacDonald}, \citenamefont {MacInnis}, \citenamefont {Macleod}, \citenamefont {Mageswaran}, \citenamefont {Mailand}, \citenamefont {M'arka}, \citenamefont {M'arka}, \citenamefont {Markosyan}, \citenamefont {Maros}, \citenamefont {Martin}, \citenamefont {Martin}, \citenamefont {Marx}, \citenamefont {Mason}, \citenamefont {Massinger}, \citenamefont {Matichard}, \citenamefont {Mavalvala}, \citenamefont {McCarthy}, \citenamefont {McClelland}, \citenamefont {McCormick}, \citenamefont {McIntyre}, \citenamefont {McIver}, \citenamefont {Merilh}, \citenamefont {Meyer}, \citenamefont {Meyers}, \citenamefont {Miller}, \citenamefont {Mittleman}, \citenamefont {Moreno}, \citenamefont {Mueller}, \citenamefont {Mueller}, \citenamefont {Mullavey}, \citenamefont {Munch}, \citenamefont {Murray}, \citenamefont {Nuttall}, \citenamefont {Oberling}, \citenamefont {O'Dell},
  \citenamefont {Oppermann}, \citenamefont {Oram}, \citenamefont {O'Reilly}, \citenamefont {Osthelder}, \citenamefont {Ottaway}, \citenamefont {Overmier}, \citenamefont {Palamos}, \citenamefont {Paris}, \citenamefont {Parker}, \citenamefont {Patrick}, \citenamefont {Pele}, \citenamefont {Penn}, \citenamefont {Phelps}, \citenamefont {Pickenpack}, \citenamefont {Piero}, \citenamefont {Pinto}, \citenamefont {Poeld}, \citenamefont {Principe}, \citenamefont {Prokhorov}, \citenamefont {Puncken}, \citenamefont {Quetschke}, \citenamefont {Quintero}, \citenamefont {Raab}, \citenamefont {Radkins}, \citenamefont {Raffai}, \citenamefont {Ramet}, \citenamefont {Reed}, \citenamefont {Reid}, \citenamefont {Reitze}, \citenamefont {Robertson}, \citenamefont {Rollins}, \citenamefont {Roma}, \citenamefont {Romie}, \citenamefont {Rowan}, \citenamefont {Ryan}, \citenamefont {Sadecki}, \citenamefont {Sanchez}, \citenamefont {Sandberg}, \citenamefont {Sannibale}, \citenamefont {Savage}, \citenamefont {Schofield}, \citenamefont
  {Schultz}, \citenamefont {Schwinberg}, \citenamefont {Sellers}, \citenamefont {Sevigny}, \citenamefont {Shaddock}, \citenamefont {Shao}, \citenamefont {Shapiro}, \citenamefont {Shawhan}, \citenamefont {Shoemaker}, \citenamefont {Sigg}, \citenamefont {Slagmolen}, \citenamefont {Smith}, \citenamefont {Smith}, \citenamefont {{Smith-Lefebvre}}, \citenamefont {Sorazu}, \citenamefont {Staley}, \citenamefont {Stein}, \citenamefont {Stochino}, \citenamefont {Strain}, \citenamefont {Taylor}, \citenamefont {Thomas}, \citenamefont {Thomas}, \citenamefont {Thorne}, \citenamefont {Thrane}, \citenamefont {Tokmakov}, \citenamefont {Torrie}, \citenamefont {Traylor}, \citenamefont {Vajente}, \citenamefont {Valdes}, \citenamefont {{van Veggel}}, \citenamefont {Vargas}, \citenamefont {Vecchio}, \citenamefont {Veitch}, \citenamefont {Venkateswara}, \citenamefont {Vo}, \citenamefont {Vorvick}, \citenamefont {Waldman}, \citenamefont {Walker}, \citenamefont {Ward}, \citenamefont {Warner}, \citenamefont {Weaver}, \citenamefont
  {Weiss}, \citenamefont {Welborn}, \citenamefont {Wessels}, \citenamefont {Wilkinson}, \citenamefont {Willems}, \citenamefont {Williams}, \citenamefont {Willke}, \citenamefont {Wilmut}, \citenamefont {Winkelmann}, \citenamefont {Wipf}, \citenamefont {Worden}, \citenamefont {Wu}, \citenamefont {Yamamoto}, \citenamefont {Yancey}, \citenamefont {Yu}, \citenamefont {Zhang}, \citenamefont {Zucker},\ and\ \citenamefont {Zweizig}}]{martynov_sensitivity_2016}%
  \BibitemOpen
  \bibfield  {author} {\bibinfo {author} {\bibfnamefont {D.~V.}\ \bibnamefont {Martynov}}, \bibinfo {author} {\bibfnamefont {E.~D.}\ \bibnamefont {Hall}}, \bibinfo {author} {\bibfnamefont {B.~P.}\ \bibnamefont {Abbott}}, \bibinfo {author} {\bibfnamefont {R.}~\bibnamefont {Abbott}}, \bibinfo {author} {\bibfnamefont {T.~D.}\ \bibnamefont {Abbott}}, \bibnamefont {et~al.},\ }\href {\doibase 10.1103/PhysRevD.93.112004} {\bibfield  {journal} {\bibinfo  {journal} {Phys. Rev. D}\ }\textbf {\bibinfo {volume} {93}},\ \bibinfo {pages} {112004} (\bibinfo {year} {2016})}\BibitemShut {NoStop}%
\bibitem [{\citenamefont {Maggiore}\ \emph {et~al.}(2020)\citenamefont {Maggiore}, \citenamefont {Broeck}, \citenamefont {Bartolo}, \citenamefont {Belgacem}, \citenamefont {Bertacca}, \citenamefont {Bizouard}, \citenamefont {Branchesi}, \citenamefont {Clesse}, \citenamefont {Foffa}, \citenamefont {{Garc{\'i}a-Bellido}}, \citenamefont {Grimm}, \citenamefont {Harms}, \citenamefont {Hinderer}, \citenamefont {Matarrese}, \citenamefont {Palomba}, \citenamefont {Peloso}, \citenamefont {Ricciardone},\ and\ \citenamefont {Sakellariadou}}]{maggiore_science_2020}%
  \BibitemOpen
  \bibfield  {author} {\bibinfo {author} {\bibfnamefont {M.}~\bibnamefont {Maggiore}}, \bibinfo {author} {\bibfnamefont {C.~V.~D.}\ \bibnamefont {Broeck}}, \bibinfo {author} {\bibfnamefont {N.}~\bibnamefont {Bartolo}}, \bibinfo {author} {\bibfnamefont {E.}~\bibnamefont {Belgacem}}, \bibinfo {author} {\bibfnamefont {D.}~\bibnamefont {Bertacca}}, \bibnamefont {et~al.},\ }\href {\doibase 10.1088/1475-7516/2020/03/050} {\bibfield  {journal} {\bibinfo  {journal} {J. Cosmol. Astropart. Phys.}\ }\textbf {\bibinfo {volume} {2020}},\ \bibinfo {pages} {050} (\bibinfo {year} {2020})}\BibitemShut {NoStop}%
\bibitem [{\citenamefont {Evans}\ \emph {et~al.}(2021)\citenamefont {Evans}, \citenamefont {Adhikari}, \citenamefont {Afle}, \citenamefont {Ballmer}, \citenamefont {Biscoveanu}, \citenamefont {Borhanian}, \citenamefont {Brown}, \citenamefont {Chen}, \citenamefont {Eisenstein}, \citenamefont {Gruson}, \citenamefont {Gupta}, \citenamefont {Hall}, \citenamefont {Huxford}, \citenamefont {Kamai}, \citenamefont {Kashyap}, \citenamefont {Kissel}, \citenamefont {Kuns}, \citenamefont {Landry}, \citenamefont {Lenon}, \citenamefont {Lovelace}, \citenamefont {McCuller}, \citenamefont {Ng}, \citenamefont {Nitz}, \citenamefont {Read}, \citenamefont {Sathyaprakash}, \citenamefont {Shoemaker}, \citenamefont {Slagmolen}, \citenamefont {Smith}, \citenamefont {Srivastava}, \citenamefont {Sun}, \citenamefont {Vitale},\ and\ \citenamefont {Weiss}}]{evans_horizon_2021}%
  \BibitemOpen
  \bibfield  {author} {\bibinfo {author} {\bibfnamefont {M.}~\bibnamefont {Evans}}, \bibinfo {author} {\bibfnamefont {R.~X.}\ \bibnamefont {Adhikari}}, \bibinfo {author} {\bibfnamefont {C.}~\bibnamefont {Afle}}, \bibinfo {author} {\bibfnamefont {S.~W.}\ \bibnamefont {Ballmer}}, \bibinfo {author} {\bibfnamefont {S.}~\bibnamefont {Biscoveanu}}, \bibnamefont {et~al.},\ }\href@noop {} {\enquote {\bibinfo {title} {A {{Horizon Study}} for {{Cosmic Explorer}}: {{Science}}, {{Observatories}}, and {{Community}}},}\ } (\bibinfo {year} {2021}),\ \Eprint {http://arxiv.org/abs/2109.09882}{arXiv:2109.09882}\BibitemShut {NoStop}%
\bibitem [{\citenamefont {{Amaro-Seoane}}\ \emph {et~al.}(2017)\citenamefont {{Amaro-Seoane}}, \citenamefont {Audley}, \citenamefont {Babak}, \citenamefont {Baker}, \citenamefont {Barausse}, \citenamefont {Bender}, \citenamefont {Berti}, \citenamefont {Binetruy}, \citenamefont {Born}, \citenamefont {Bortoluzzi}, \citenamefont {Camp}, \citenamefont {Caprini}, \citenamefont {Cardoso}, \citenamefont {Colpi}, \citenamefont {Conklin}, \citenamefont {Cornish}, \citenamefont {Cutler}, \citenamefont {Danzmann}, \citenamefont {Dolesi}, \citenamefont {Ferraioli}, \citenamefont {Ferroni}, \citenamefont {Fitzsimons}, \citenamefont {Gair}, \citenamefont {Bote}, \citenamefont {Giardini}, \citenamefont {Gibert}, \citenamefont {Grimani}, \citenamefont {Halloin}, \citenamefont {Heinzel}, \citenamefont {Hertog}, \citenamefont {Hewitson}, \citenamefont {{Holley-Bockelmann}}, \citenamefont {Hollington}, \citenamefont {Hueller}, \citenamefont {Inchauspe}, \citenamefont {Jetzer}, \citenamefont {Karnesis}, \citenamefont {Killow},
  \citenamefont {Klein}, \citenamefont {Klipstein}, \citenamefont {Korsakova}, \citenamefont {Larson}, \citenamefont {Livas}, \citenamefont {Lloro}, \citenamefont {Man}, \citenamefont {Mance}, \citenamefont {Martino}, \citenamefont {Mateos}, \citenamefont {McKenzie}, \citenamefont {McWilliams}, \citenamefont {Miller}, \citenamefont {Mueller}, \citenamefont {Nardini}, \citenamefont {Nelemans}, \citenamefont {Nofrarias}, \citenamefont {Petiteau}, \citenamefont {Pivato}, \citenamefont {Plagnol}, \citenamefont {Porter}, \citenamefont {Reiche}, \citenamefont {Robertson}, \citenamefont {Robertson}, \citenamefont {Rossi}, \citenamefont {Russano}, \citenamefont {Schutz}, \citenamefont {Sesana}, \citenamefont {Shoemaker}, \citenamefont {Slutsky}, \citenamefont {Sopuerta}, \citenamefont {Sumner}, \citenamefont {Tamanini}, \citenamefont {Thorpe}, \citenamefont {Troebs}, \citenamefont {Vallisneri}, \citenamefont {Vecchio}, \citenamefont {Vetrugno}, \citenamefont {Vitale}, \citenamefont {Volonteri}, \citenamefont
  {Wanner}, \citenamefont {Ward}, \citenamefont {Wass}, \citenamefont {Weber}, \citenamefont {Ziemer},\ and\ \citenamefont {Zweifel}}]{amaro-seoane_laser_2017}%
  \BibitemOpen
  \bibfield  {author} {\bibinfo {author} {\bibfnamefont {P.}~\bibnamefont {{Amaro-Seoane}}}, \bibinfo {author} {\bibfnamefont {H.}~\bibnamefont {Audley}}, \bibinfo {author} {\bibfnamefont {S.}~\bibnamefont {Babak}}, \bibinfo {author} {\bibfnamefont {J.}~\bibnamefont {Baker}}, \bibinfo {author} {\bibfnamefont {E.}~\bibnamefont {Barausse}}, \bibnamefont {et~al.},\ }\href@noop {} {\enquote {\bibinfo {title} {Laser {{Interferometer Space Antenna}}},}\ } (\bibinfo {year} {2017}),\ \Eprint {http://arxiv.org/abs/1702.00786}{arXiv:1702.00786}\BibitemShut {NoStop}%
\bibitem [{\citenamefont {Hu}\ and\ \citenamefont {Wu}(2017)}]{hu_taiji_2017}%
  \BibitemOpen
  \bibfield  {author} {\bibinfo {author} {\bibfnamefont {W.-R.}\ \bibnamefont {Hu}}\ \bibnamefont {and}\ \bibinfo {author} {\bibfnamefont {Y.-L.}\ \bibnamefont {Wu}},\ }\href {\doibase 10.1093/nsr/nwx116} {\bibfield  {journal} {\bibinfo  {journal} {Natl. Sci. Rev.}\ }\textbf {\bibinfo {volume} {4}},\ \bibinfo {pages} {685} (\bibinfo {year} {2017})}\BibitemShut {NoStop}%
\bibitem [{\citenamefont {Ren}\ \emph {et~al.}(2023)\citenamefont {Ren}, \citenamefont {Zhao}, \citenamefont {Cao}, \citenamefont {Guo}, \citenamefont {Han}, \citenamefont {Jin},\ and\ \citenamefont {Wu}}]{ren_taiji_2023}%
  \BibitemOpen
  \bibfield  {author} {\bibinfo {author} {\bibfnamefont {Z.}~\bibnamefont {Ren}}, \bibinfo {author} {\bibfnamefont {T.}~\bibnamefont {Zhao}}, \bibinfo {author} {\bibfnamefont {Z.}~\bibnamefont {Cao}}, \bibinfo {author} {\bibfnamefont {Z.-K.}\ \bibnamefont {Guo}}, \bibinfo {author} {\bibfnamefont {W.-B.}\ \bibnamefont {Han}}, \bibnamefont {et~al.},\ }\href {\doibase 10.1007/s11467-023-1318-y} {\bibfield  {journal} {\bibinfo  {journal} {Front. Phys.}\ }\textbf {\bibinfo {volume} {18}},\ \bibinfo {pages} {64302} (\bibinfo {year} {2023})}\BibitemShut {NoStop}%
\bibitem [{\citenamefont {Luo}\ \emph {et~al.}(2016)\citenamefont {Luo}, \citenamefont {Chen}, \citenamefont {Duan}, \citenamefont {Gong}, \citenamefont {Hu}, \citenamefont {Ji}, \citenamefont {Liu}, \citenamefont {Mei}, \citenamefont {Milyukov}, \citenamefont {Sazhin}, \citenamefont {Shao}, \citenamefont {Toth}, \citenamefont {Tu}, \citenamefont {Wang}, \citenamefont {Wang}, \citenamefont {Yeh}, \citenamefont {Zhan}, \citenamefont {Zhang}, \citenamefont {Zharov},\ and\ \citenamefont {Zhou}}]{luo_tianqin_2016}%
  \BibitemOpen
  \bibfield  {author} {\bibinfo {author} {\bibfnamefont {J.}~\bibnamefont {Luo}}, \bibinfo {author} {\bibfnamefont {L.-S.}\ \bibnamefont {Chen}}, \bibinfo {author} {\bibfnamefont {H.-Z.}\ \bibnamefont {Duan}}, \bibinfo {author} {\bibfnamefont {Y.-G.}\ \bibnamefont {Gong}}, \bibinfo {author} {\bibfnamefont {S.}~\bibnamefont {Hu}}, \bibnamefont {et~al.},\ }\href {\doibase 10.1088/0264-9381/33/3/035010} {\bibfield  {journal} {\bibinfo  {journal} {Class. Quantum Gravity}\ }\textbf {\bibinfo {volume} {33}},\ \bibinfo {pages} {035010} (\bibinfo {year} {2016})},\ \Eprint {http://arxiv.org/abs/1512.02076}{arXiv:1512.02076}\BibitemShut {NoStop}%
\bibitem [{\citenamefont {Bailes}\ \emph {et~al.}(2021)\citenamefont {Bailes}, \citenamefont {Berger}, \citenamefont {Brady}, \citenamefont {Branchesi}, \citenamefont {Danzmann}, \citenamefont {Evans}, \citenamefont {{Holley-Bockelmann}}, \citenamefont {Iyer}, \citenamefont {Kajita}, \citenamefont {Katsanevas}, \citenamefont {Kramer}, \citenamefont {Lazzarini}, \citenamefont {Lehner}, \citenamefont {Losurdo}, \citenamefont {L{\"u}ck}, \citenamefont {McClelland}, \citenamefont {McLaughlin}, \citenamefont {Punturo}, \citenamefont {Ransom}, \citenamefont {Raychaudhury}, \citenamefont {Reitze}, \citenamefont {Ricci}, \citenamefont {Rowan}, \citenamefont {Saito}, \citenamefont {Sanders}, \citenamefont {Sathyaprakash}, \citenamefont {Schutz}, \citenamefont {Sesana}, \citenamefont {Shinkai}, \citenamefont {Siemens}, \citenamefont {Shoemaker}, \citenamefont {Thorpe}, \citenamefont {{van den Brand}},\ and\ \citenamefont {Vitale}}]{bailes_gravitational-wave_2021}%
  \BibitemOpen
  \bibfield  {author} {\bibinfo {author} {\bibfnamefont {M.}~\bibnamefont {Bailes}}, \bibinfo {author} {\bibfnamefont {B.~K.}\ \bibnamefont {Berger}}, \bibinfo {author} {\bibfnamefont {P.~R.}\ \bibnamefont {Brady}}, \bibinfo {author} {\bibfnamefont {M.}~\bibnamefont {Branchesi}}, \bibinfo {author} {\bibfnamefont {K.}~\bibnamefont {Danzmann}}, \bibnamefont {et~al.},\ }\href {\doibase 10.1038/s42254-021-00303-8} {\bibfield  {journal} {\bibinfo  {journal} {Nat. Rev. Phys.}\ }\textbf {\bibinfo {volume} {3}},\ \bibinfo {pages} {344} (\bibinfo {year} {2021})}\BibitemShut {NoStop}%
\bibitem [{\citenamefont {Wang}\ \emph {et~al.}(2025)\citenamefont {Wang}, \citenamefont {Hu},\ and\ \citenamefont {Shao}}]{wang_rigorous_2025}%
  \BibitemOpen
  \bibfield  {author} {\bibinfo {author} {\bibfnamefont {Z.}~\bibnamefont {Wang}}, \bibinfo {author} {\bibfnamefont {Z.}~\bibnamefont {Hu}},\ \bibnamefont {and}\ \bibinfo {author} {\bibfnamefont {L.}~\bibnamefont {Shao}},\ }\href@noop {} {\enquote {\bibinfo {title} {Rigorous analytic solution to the gravitational-wave overlapping event rates},}\ } (\bibinfo {year} {2025}),\ \Eprint {http://arxiv.org/abs/2501.05218}{arXiv:2501.05218}\BibitemShut {NoStop}%
\bibitem [{\citenamefont {Iacovelli}\ \emph {et~al.}(2022)\citenamefont {Iacovelli}, \citenamefont {Mancarella}, \citenamefont {Foffa},\ and\ \citenamefont {Maggiore}}]{iacovelli_forecasting_2022}%
  \BibitemOpen
  \bibfield  {author} {\bibinfo {author} {\bibfnamefont {F.}~\bibnamefont {Iacovelli}}, \bibinfo {author} {\bibfnamefont {M.}~\bibnamefont {Mancarella}}, \bibinfo {author} {\bibfnamefont {S.}~\bibnamefont {Foffa}},\ \bibnamefont {and}\ \bibinfo {author} {\bibfnamefont {M.}~\bibnamefont {Maggiore}},\ }\href {\doibase 10.3847/1538-4357/ac9cd4} {\bibfield  {journal} {\bibinfo  {journal} {Astrophys. J.}\ }\textbf {\bibinfo {volume} {941}},\ \bibinfo {pages} {208} (\bibinfo {year} {2022})}\BibitemShut {NoStop}%
\bibitem [{\citenamefont {Pieroni}\ \emph {et~al.}(2022)\citenamefont {Pieroni}, \citenamefont {Ricciardone},\ and\ \citenamefont {Barausse}}]{pieroni_detectability_2022}%
  \BibitemOpen
  \bibfield  {author} {\bibinfo {author} {\bibfnamefont {M.}~\bibnamefont {Pieroni}}, \bibinfo {author} {\bibfnamefont {A.}~\bibnamefont {Ricciardone}},\ \bibnamefont {and}\ \bibinfo {author} {\bibfnamefont {E.}~\bibnamefont {Barausse}},\ }\href {\doibase 10.1038/s41598-022-19540-7} {\bibfield  {journal} {\bibinfo  {journal} {Sci. Rep.}\ }\textbf {\bibinfo {volume} {12}},\ \bibinfo {pages} {17940} (\bibinfo {year} {2022})}\BibitemShut {NoStop}%
\bibitem [{\citenamefont {Digman}\ and\ \citenamefont {Cornish}(2022)}]{digman_lisa_2022}%
  \BibitemOpen
  \bibfield  {author} {\bibinfo {author} {\bibfnamefont {M.~C.}\ \bibnamefont {Digman}}\ \bibnamefont {and}\ \bibinfo {author} {\bibfnamefont {N.~J.}\ \bibnamefont {Cornish}},\ }\href {\doibase 10.3847/1538-4357/ac9139} {\bibfield  {journal} {\bibinfo  {journal} {Astrophys. J.}\ }\textbf {\bibinfo {volume} {940}},\ \bibinfo {pages} {10} (\bibinfo {year} {2022})}\BibitemShut {NoStop}%
\bibitem [{\citenamefont {Arca~Sedda}\ \emph {et~al.}(2021)\citenamefont {Arca~Sedda}, \citenamefont {Amaro~Seoane},\ and\ \citenamefont {Chen}}]{arca_sedda_merging_2021}%
  \BibitemOpen
  \bibfield  {author} {\bibinfo {author} {\bibfnamefont {M.}~\bibnamefont {Arca~Sedda}}, \bibinfo {author} {\bibfnamefont {P.}~\bibnamefont {Amaro~Seoane}},\ \bibnamefont {and}\ \bibinfo {author} {\bibfnamefont {X.}~\bibnamefont {Chen}},\ }\href {\doibase 10.1051/0004-6361/202037785} {\bibfield  {journal} {\bibinfo  {journal} {Astron. Astrophys.}\ }\textbf {\bibinfo {volume} {652}},\ \bibinfo {pages} {A54} (\bibinfo {year} {2021})}\BibitemShut {NoStop}%
\bibitem [{\citenamefont {Baghi}(2022)}]{baghi_2022}%
  \BibitemOpen
  \bibfield  {author} {\bibinfo {author} {\bibfnamefont {Q.}~\bibnamefont {Baghi}},\ }\href@noop {} {\enquote {\bibinfo {title} {The {{LISA Data Challenges}}},}\ } (\bibinfo {year} {2022}),\ \Eprint {http://arxiv.org/abs/2204.12142}{arXiv:2204.12142}\BibitemShut {NoStop}%
\bibitem [{\citenamefont {Himemoto}\ \emph {et~al.}(2021)\citenamefont {Himemoto}, \citenamefont {Nishizawa},\ and\ \citenamefont {Taruya}}]{himemoto_impacts_2021}%
  \BibitemOpen
  \bibfield  {author} {\bibinfo {author} {\bibfnamefont {Y.}~\bibnamefont {Himemoto}}, \bibinfo {author} {\bibfnamefont {A.}~\bibnamefont {Nishizawa}},\ \bibnamefont {and}\ \bibinfo {author} {\bibfnamefont {A.}~\bibnamefont {Taruya}},\ }\href {\doibase 10.1103/PhysRevD.104.044010} {\bibfield  {journal} {\bibinfo  {journal} {Phys. Rev. D}\ }\textbf {\bibinfo {volume} {104}},\ \bibinfo {pages} {044010} (\bibinfo {year} {2021})}\BibitemShut {NoStop}%
\bibitem [{\citenamefont {Chamberlain}\ and\ \citenamefont {Yunes}(2017)}]{chamberlain_theoretical_2017}%
  \BibitemOpen
  \bibfield  {author} {\bibinfo {author} {\bibfnamefont {K.}~\bibnamefont {Chamberlain}}\ \bibnamefont {and}\ \bibinfo {author} {\bibfnamefont {N.}~\bibnamefont {Yunes}},\ }\href {\doibase 10.1103/PhysRevD.96.084039} {\bibfield  {journal} {\bibinfo  {journal} {Phys. Rev. D}\ }\textbf {\bibinfo {volume} {96}},\ \bibinfo {pages} {084039} (\bibinfo {year} {2017})}\BibitemShut {NoStop}%
\bibitem [{\citenamefont {Johnson}\ \emph {et~al.}(2024)\citenamefont {Johnson}, \citenamefont {Chatziioannou},\ and\ \citenamefont {Farr}}]{johnson_source_2024}%
  \BibitemOpen
  \bibfield  {author} {\bibinfo {author} {\bibfnamefont {A.~D.}\ \bibnamefont {Johnson}}, \bibinfo {author} {\bibfnamefont {K.}~\bibnamefont {Chatziioannou}},\ \bibnamefont {and}\ \bibinfo {author} {\bibfnamefont {W.~M.}\ \bibnamefont {Farr}},\ }\href {\doibase 10.1103/PhysRevD.109.084015} {\bibfield  {journal} {\bibinfo  {journal} {Phys. Rev. D}\ }\textbf {\bibinfo {volume} {109}},\ \bibinfo {pages} {084015} (\bibinfo {year} {2024})}\BibitemShut {NoStop}%
\bibitem [{\citenamefont {Hourihane}\ \emph {et~al.}(2022)\citenamefont {Hourihane}, \citenamefont {Chatziioannou}, \citenamefont {Wijngaarden}, \citenamefont {Davis}, \citenamefont {Littenberg},\ and\ \citenamefont {Cornish}}]{hourihane_accurate_2022}%
  \BibitemOpen
  \bibfield  {author} {\bibinfo {author} {\bibfnamefont {S.}~\bibnamefont {Hourihane}}, \bibinfo {author} {\bibfnamefont {K.}~\bibnamefont {Chatziioannou}}, \bibinfo {author} {\bibfnamefont {M.}~\bibnamefont {Wijngaarden}}, \bibinfo {author} {\bibfnamefont {D.}~\bibnamefont {Davis}}, \bibinfo {author} {\bibfnamefont {T.}~\bibnamefont {Littenberg}}, \bibnamefont {et~al.},\ }\href {\doibase 10.1103/PhysRevD.106.042006} {\bibfield  {journal} {\bibinfo  {journal} {Phys. Rev. D}\ }\textbf {\bibinfo {volume} {106}},\ \bibinfo {pages} {042006} (\bibinfo {year} {2022})}\BibitemShut {NoStop}%
\bibitem [{\citenamefont {Regimbau}\ and\ \citenamefont {Hughes}(2009)}]{regimbau_gravitational-wave_2009}%
  \BibitemOpen
  \bibfield  {author} {\bibinfo {author} {\bibfnamefont {T.}~\bibnamefont {Regimbau}}\ \bibnamefont {and}\ \bibinfo {author} {\bibfnamefont {S.~A.}\ \bibnamefont {Hughes}},\ }\href {\doibase 10.1103/PhysRevD.79.062002} {\bibfield  {journal} {\bibinfo  {journal} {Phys. Rev. D}\ }\textbf {\bibinfo {volume} {79}},\ \bibinfo {pages} {062002} (\bibinfo {year} {2009})}\BibitemShut {NoStop}%
\bibitem [{\citenamefont {Wu}\ and\ \citenamefont {Nitz}(2023)}]{wu_mock_2023}%
  \BibitemOpen
  \bibfield  {author} {\bibinfo {author} {\bibfnamefont {S.}~\bibnamefont {Wu}}\ \bibnamefont {and}\ \bibinfo {author} {\bibfnamefont {A.~H.}\ \bibnamefont {Nitz}},\ }\href {\doibase 10.1103/PhysRevD.107.063022} {\bibfield  {journal} {\bibinfo  {journal} {Phys. Rev. D}\ }\textbf {\bibinfo {volume} {107}},\ \bibinfo {pages} {063022} (\bibinfo {year} {2023})}\BibitemShut {NoStop}%
\bibitem [{\citenamefont {Badaracco}\ \emph {et~al.}(2024)\citenamefont {Badaracco}, \citenamefont {Banerjee}, \citenamefont {Branchesi},\ and\ \citenamefont {Chincarini}}]{badaracco_blind_2024}%
  \BibitemOpen
  \bibfield  {author} {\bibinfo {author} {\bibfnamefont {F.}~\bibnamefont {Badaracco}}, \bibinfo {author} {\bibfnamefont {B.}~\bibnamefont {Banerjee}}, \bibinfo {author} {\bibfnamefont {M.}~\bibnamefont {Branchesi}},\ \bibnamefont {and}\ \bibinfo {author} {\bibfnamefont {A.}~\bibnamefont {Chincarini}},\ }\href {\doibase 10.1016/j.newar.2024.101707} {\bibfield  {journal} {\bibinfo  {journal} {New Astron. Rev.}\ }\textbf {\bibinfo {volume} {99}},\ \bibinfo {pages} {101707} (\bibinfo {year} {2024})}\BibitemShut {NoStop}%
\bibitem [{\citenamefont {Relton}\ and\ \citenamefont {Raymond}(2021)}]{relton_parameter_2021}%
  \BibitemOpen
  \bibfield  {author} {\bibinfo {author} {\bibfnamefont {P.}~\bibnamefont {Relton}}\ \bibnamefont {and}\ \bibinfo {author} {\bibfnamefont {V.}~\bibnamefont {Raymond}},\ }\href {\doibase 10.1103/PhysRevD.104.084039} {\bibfield  {journal} {\bibinfo  {journal} {Phys. Rev. D}\ }\textbf {\bibinfo {volume} {104}},\ \bibinfo {pages} {084039} (\bibinfo {year} {2021})}\BibitemShut {NoStop}%
\bibitem [{\citenamefont {Samajdar}\ \emph {et~al.}(2021)\citenamefont {Samajdar}, \citenamefont {Janquart}, \citenamefont {Van Den~Broeck},\ and\ \citenamefont {Dietrich}}]{samajdar_biases_2021}%
  \BibitemOpen
  \bibfield  {author} {\bibinfo {author} {\bibfnamefont {A.}~\bibnamefont {Samajdar}}, \bibinfo {author} {\bibfnamefont {J.}~\bibnamefont {Janquart}}, \bibinfo {author} {\bibfnamefont {C.}~\bibnamefont {Van Den~Broeck}},\ \bibnamefont {and}\ \bibinfo {author} {\bibfnamefont {T.}~\bibnamefont {Dietrich}},\ }\href {\doibase 10.1103/PhysRevD.104.044003} {\bibfield  {journal} {\bibinfo  {journal} {Phys. Rev. D}\ }\textbf {\bibinfo {volume} {104}},\ \bibinfo {pages} {044003} (\bibinfo {year} {2021})}\BibitemShut {NoStop}%
\bibitem [{\citenamefont {Pizzati}\ \emph {et~al.}(2022)\citenamefont {Pizzati}, \citenamefont {Sachdev}, \citenamefont {Gupta},\ and\ \citenamefont {Sathyaprakash}}]{pizzati_toward_2022}%
  \BibitemOpen
  \bibfield  {author} {\bibinfo {author} {\bibfnamefont {E.}~\bibnamefont {Pizzati}}, \bibinfo {author} {\bibfnamefont {S.}~\bibnamefont {Sachdev}}, \bibinfo {author} {\bibfnamefont {A.}~\bibnamefont {Gupta}},\ \bibnamefont {and}\ \bibinfo {author} {\bibfnamefont {B.~S.}\ \bibnamefont {Sathyaprakash}},\ }\href {\doibase 10.1103/PhysRevD.105.104016} {\bibfield  {journal} {\bibinfo  {journal} {Phys. Rev. D}\ }\textbf {\bibinfo {volume} {105}},\ \bibinfo {pages} {104016} (\bibinfo {year} {2022})}\BibitemShut {NoStop}%
\bibitem [{\citenamefont {Janquart}\ \emph {et~al.}(2023)\citenamefont {Janquart}, \citenamefont {Baka}, \citenamefont {Samajdar}, \citenamefont {Dietrich},\ and\ \citenamefont {Van~Den~Broeck}}]{janquart_analyses_2023}%
  \BibitemOpen
  \bibfield  {author} {\bibinfo {author} {\bibfnamefont {J.}~\bibnamefont {Janquart}}, \bibinfo {author} {\bibfnamefont {T.}~\bibnamefont {Baka}}, \bibinfo {author} {\bibfnamefont {A.}~\bibnamefont {Samajdar}}, \bibinfo {author} {\bibfnamefont {T.}~\bibnamefont {Dietrich}},\ \bibnamefont {and}\ \bibinfo {author} {\bibfnamefont {C.}~\bibnamefont {Van~Den~Broeck}},\ }\href {\doibase 10.1093/mnras/stad1542} {\bibfield  {journal} {\bibinfo  {journal} {Mon. Not. R. Astron. Soc.}\ }\textbf {\bibinfo {volume} {523}},\ \bibinfo {pages} {1699} (\bibinfo {year} {2023})}\BibitemShut {NoStop}%
\bibitem [{\citenamefont {Antonelli}\ \emph {et~al.}(2021)\citenamefont {Antonelli}, \citenamefont {Burke},\ and\ \citenamefont {Gair}}]{antonelli_noisy_2021}%
  \BibitemOpen
  \bibfield  {author} {\bibinfo {author} {\bibfnamefont {A.}~\bibnamefont {Antonelli}}, \bibinfo {author} {\bibfnamefont {O.}~\bibnamefont {Burke}},\ \bibnamefont {and}\ \bibinfo {author} {\bibfnamefont {J.~R.}\ \bibnamefont {Gair}},\ }\href {\doibase 10.1093/mnras/stab2358} {\bibfield  {journal} {\bibinfo  {journal} {Mon. Not. R. Astron. Soc.}\ }\textbf {\bibinfo {volume} {507}},\ \bibinfo {pages} {5069} (\bibinfo {year} {2021})}\BibitemShut {NoStop}%
\bibitem [{\citenamefont {Wang}\ \emph {et~al.}(2024{\natexlab{a}})\citenamefont {Wang}, \citenamefont {Liang}, \citenamefont {Zhao}, \citenamefont {Liu},\ and\ \citenamefont {Shao}}]{wang_anatomy_2024}%
  \BibitemOpen
  \bibfield  {author} {\bibinfo {author} {\bibfnamefont {Z.}~\bibnamefont {Wang}}, \bibinfo {author} {\bibfnamefont {D.}~\bibnamefont {Liang}}, \bibinfo {author} {\bibfnamefont {J.}~\bibnamefont {Zhao}}, \bibinfo {author} {\bibfnamefont {C.}~\bibnamefont {Liu}},\ \bibnamefont {and}\ \bibinfo {author} {\bibfnamefont {L.}~\bibnamefont {Shao}},\ }\href {\doibase 10.1088/1361-6382/ad210b} {\bibfield  {journal} {\bibinfo  {journal} {Class. Quantum Gravity}\ }\textbf {\bibinfo {volume} {41}},\ \bibinfo {pages} {055011} (\bibinfo {year} {2024}{\natexlab{a}})}\BibitemShut {NoStop}%
\bibitem [{\citenamefont {Relton}\ \emph {et~al.}(2022)\citenamefont {Relton}, \citenamefont {Virtuoso}, \citenamefont {Bini}, \citenamefont {Raymond}, \citenamefont {Harry}, \citenamefont {Drago}, \citenamefont {Lazzaro}, \citenamefont {Miani},\ and\ \citenamefont {Tiwari}}]{relton_addressing_2022}%
  \BibitemOpen
  \bibfield  {author} {\bibinfo {author} {\bibfnamefont {P.}~\bibnamefont {Relton}}, \bibinfo {author} {\bibfnamefont {A.}~\bibnamefont {Virtuoso}}, \bibinfo {author} {\bibfnamefont {S.}~\bibnamefont {Bini}}, \bibinfo {author} {\bibfnamefont {V.}~\bibnamefont {Raymond}}, \bibinfo {author} {\bibfnamefont {I.}~\bibnamefont {Harry}}, \bibnamefont {et~al.},\ }\href {\doibase 10.1103/PhysRevD.106.104045} {\bibfield  {journal} {\bibinfo  {journal} {Phys. Rev. D}\ }\textbf {\bibinfo {volume} {106}},\ \bibinfo {pages} {104045} (\bibinfo {year} {2022})}\BibitemShut {NoStop}%
\bibitem [{\citenamefont {Kerachian}\ \emph {et~al.}(2024)\citenamefont {Kerachian}, \citenamefont {Mukherjee}, \citenamefont {{Lukes-Gerakopoulos}},\ and\ \citenamefont {Mitra}}]{kerachian_detectability_2024}%
  \BibitemOpen
  \bibfield  {author} {\bibinfo {author} {\bibfnamefont {M.}~\bibnamefont {Kerachian}}, \bibinfo {author} {\bibfnamefont {S.}~\bibnamefont {Mukherjee}}, \bibinfo {author} {\bibfnamefont {G.}~\bibnamefont {{Lukes-Gerakopoulos}}},\ \bibnamefont {and}\ \bibinfo {author} {\bibfnamefont {S.}~\bibnamefont {Mitra}},\ }\href {\doibase 10.1051/0004-6361/202348747} {\bibfield  {journal} {\bibinfo  {journal} {Astron. Astrophys.}\ }\textbf {\bibinfo {volume} {684}},\ \bibinfo {pages} {A17} (\bibinfo {year} {2024})}\BibitemShut {NoStop}%
\bibitem [{\citenamefont {Hu}\ and\ \citenamefont {Veitch}(2023)}]{hu_accumulating_2023}%
  \BibitemOpen
  \bibfield  {author} {\bibinfo {author} {\bibfnamefont {Q.}~\bibnamefont {Hu}}\ \bibnamefont {and}\ \bibinfo {author} {\bibfnamefont {J.}~\bibnamefont {Veitch}},\ }\href {\doibase 10.3847/1538-4357/acbc18} {\bibfield  {journal} {\bibinfo  {journal} {Astrophys. J.}\ }\textbf {\bibinfo {volume} {945}},\ \bibinfo {pages} {103} (\bibinfo {year} {2023})}\BibitemShut {NoStop}%
\bibitem [{\citenamefont {Dang}\ \emph {et~al.}(2024)\citenamefont {Dang}, \citenamefont {Wang}, \citenamefont {Liang},\ and\ \citenamefont {Shao}}]{dang_impact_2024}%
  \BibitemOpen
  \bibfield  {author} {\bibinfo {author} {\bibfnamefont {Y.}~\bibnamefont {Dang}}, \bibinfo {author} {\bibfnamefont {Z.}~\bibnamefont {Wang}}, \bibinfo {author} {\bibfnamefont {D.}~\bibnamefont {Liang}},\ \bibnamefont {and}\ \bibinfo {author} {\bibfnamefont {L.}~\bibnamefont {Shao}},\ }\href {\doibase 10.3847/1538-4357/ad2e00} {\bibfield  {journal} {\bibinfo  {journal} {Astrophys. J.}\ }\textbf {\bibinfo {volume} {964}},\ \bibinfo {pages} {194} (\bibinfo {year} {2024})}\BibitemShut {NoStop}%
\bibitem [{\citenamefont {Zhao}\ \emph {et~al.}(2023{\natexlab{a}})\citenamefont {Zhao}, \citenamefont {Shi}, \citenamefont {Zhou}, \citenamefont {Cao},\ and\ \citenamefont {Ren}}]{zhao_dawn_2023}%
  \BibitemOpen
  \bibfield  {author} {\bibinfo {author} {\bibfnamefont {T.}~\bibnamefont {Zhao}}, \bibinfo {author} {\bibfnamefont {R.}~\bibnamefont {Shi}}, \bibinfo {author} {\bibfnamefont {Y.}~\bibnamefont {Zhou}}, \bibinfo {author} {\bibfnamefont {Z.}~\bibnamefont {Cao}},\ \bibnamefont {and}\ \bibinfo {author} {\bibfnamefont {Z.}~\bibnamefont {Ren}},\ }\href@noop {} {\enquote {\bibinfo {title} {Dawning of a new era in gravitational wave data analysis: Unveiling cosmic mysteries via artificial intelligence -- a systematic review},}\ } (\bibinfo {year} {2023}{\natexlab{a}}),\ \Eprint {http://arxiv.org/abs/2311.15585}{arXiv:2311.15585}\BibitemShut {NoStop}%
\bibitem [{\citenamefont {Shi}\ \emph {et~al.}(2024)\citenamefont {Shi}, \citenamefont {Zhou}, \citenamefont {Zhao}, \citenamefont {Cao},\ and\ \citenamefont {Ren}}]{shi_compact_2024}%
  \BibitemOpen
  \bibfield  {author} {\bibinfo {author} {\bibfnamefont {R.}~\bibnamefont {Shi}}, \bibinfo {author} {\bibfnamefont {Y.}~\bibnamefont {Zhou}}, \bibinfo {author} {\bibfnamefont {T.}~\bibnamefont {Zhao}}, \bibinfo {author} {\bibfnamefont {Z.}~\bibnamefont {Cao}},\ \bibnamefont {and}\ \bibinfo {author} {\bibfnamefont {Z.}~\bibnamefont {Ren}},\ }\href {\doibase 10.1103/PhysRevD.109.084017} {\bibfield  {journal} {\bibinfo  {journal} {Phys. Rev. D}\ }\textbf {\bibinfo {volume} {109}},\ \bibinfo {pages} {084017} (\bibinfo {year} {2024})}\BibitemShut {NoStop}%
\bibitem [{\citenamefont {Shi}\ \emph {et~al.}(2025)\citenamefont {Shi}, \citenamefont {Zhou}, \citenamefont {Zhao}, \citenamefont {Wang}, \citenamefont {Ren},\ and\ \citenamefont {Cao}}]{shi_rapid_2025}%
  \BibitemOpen
  \bibfield  {author} {\bibinfo {author} {\bibfnamefont {R.}~\bibnamefont {Shi}}, \bibinfo {author} {\bibfnamefont {Y.}~\bibnamefont {Zhou}}, \bibinfo {author} {\bibfnamefont {T.}~\bibnamefont {Zhao}}, \bibinfo {author} {\bibfnamefont {Z.}~\bibnamefont {Wang}}, \bibinfo {author} {\bibfnamefont {Z.}~\bibnamefont {Ren}}, \bibnamefont {et~al.},\ }\href {\doibase 10.1103/PhysRevD.111.044016} {\bibfield  {journal} {\bibinfo  {journal} {Phys. Rev. D}\ }\textbf {\bibinfo {volume} {111}},\ \bibinfo {pages} {044016} (\bibinfo {year} {2025})}\BibitemShut {NoStop}%
\bibitem [{\citenamefont {George}\ and\ \citenamefont {Huerta}(2018)}]{george_deep_2018}%
  \BibitemOpen
  \bibfield  {author} {\bibinfo {author} {\bibfnamefont {D.}~\bibnamefont {George}}\ \bibnamefont {and}\ \bibinfo {author} {\bibfnamefont {E.~A.}\ \bibnamefont {Huerta}},\ }\href {\doibase 10.1103/PhysRevD.97.044039} {\bibfield  {journal} {\bibinfo  {journal} {Phys. Rev. D}\ }\textbf {\bibinfo {volume} {97}},\ \bibinfo {pages} {044039} (\bibinfo {year} {2018})}\BibitemShut {NoStop}%
\bibitem [{\citenamefont {Gabbard}\ \emph {et~al.}(2018)\citenamefont {Gabbard}, \citenamefont {Williams}, \citenamefont {Hayes},\ and\ \citenamefont {Messenger}}]{gabbard_matching_2018}%
  \BibitemOpen
  \bibfield  {author} {\bibinfo {author} {\bibfnamefont {H.}~\bibnamefont {Gabbard}}, \bibinfo {author} {\bibfnamefont {M.}~\bibnamefont {Williams}}, \bibinfo {author} {\bibfnamefont {F.}~\bibnamefont {Hayes}},\ \bibnamefont {and}\ \bibinfo {author} {\bibfnamefont {C.}~\bibnamefont {Messenger}},\ }\href {\doibase 10.1103/PhysRevLett.120.141103} {\bibfield  {journal} {\bibinfo  {journal} {Phys. Rev. Lett.}\ }\textbf {\bibinfo {volume} {120}},\ \bibinfo {pages} {141103} (\bibinfo {year} {2018})}\BibitemShut {NoStop}%
\bibitem [{\citenamefont {Zhao}\ \emph {et~al.}(2024)\citenamefont {Zhao}, \citenamefont {Zhou}, \citenamefont {Shi}, \citenamefont {Cao},\ and\ \citenamefont {Ren}}]{zhao_dilated_2024}%
  \BibitemOpen
  \bibfield  {author} {\bibinfo {author} {\bibfnamefont {T.}~\bibnamefont {Zhao}}, \bibinfo {author} {\bibfnamefont {Y.}~\bibnamefont {Zhou}}, \bibinfo {author} {\bibfnamefont {R.}~\bibnamefont {Shi}}, \bibinfo {author} {\bibfnamefont {Z.}~\bibnamefont {Cao}},\ \bibnamefont {and}\ \bibinfo {author} {\bibfnamefont {Z.}~\bibnamefont {Ren}},\ }\href {\doibase 10.1103/PhysRevD.109.084054} {\bibfield  {journal} {\bibinfo  {journal} {Phys. Rev. D}\ }\textbf {\bibinfo {volume} {109}},\ \bibinfo {pages} {084054} (\bibinfo {year} {2024})}\BibitemShut {NoStop}%
\bibitem [{\citenamefont {Huerta}\ \emph {et~al.}(2021)\citenamefont {Huerta}, \citenamefont {Khan}, \citenamefont {Huang}, \citenamefont {Tian}, \citenamefont {Levental}, \citenamefont {Chard}, \citenamefont {Wei}, \citenamefont {Heflin}, \citenamefont {Katz}, \citenamefont {Kindratenko}, \citenamefont {Mu}, \citenamefont {Blaiszik},\ and\ \citenamefont {Foster}}]{huerta_accelerated_2021}%
  \BibitemOpen
  \bibfield  {author} {\bibinfo {author} {\bibfnamefont {E.~A.}\ \bibnamefont {Huerta}}, \bibinfo {author} {\bibfnamefont {A.}~\bibnamefont {Khan}}, \bibinfo {author} {\bibfnamefont {X.}~\bibnamefont {Huang}}, \bibinfo {author} {\bibfnamefont {M.}~\bibnamefont {Tian}}, \bibinfo {author} {\bibfnamefont {M.}~\bibnamefont {Levental}}, \bibnamefont {et~al.},\ }\href {\doibase 10.1038/s41550-021-01405-0} {\bibfield  {journal} {\bibinfo  {journal} {Nat. Astron.}\ }\textbf {\bibinfo {volume} {5}},\ \bibinfo {pages} {1062} (\bibinfo {year} {2021})}\BibitemShut {NoStop}%
\bibitem [{\citenamefont {Zhao}\ \emph {et~al.}(2023{\natexlab{b}})\citenamefont {Zhao}, \citenamefont {Lyu}, \citenamefont {Wang}, \citenamefont {Cao},\ and\ \citenamefont {Ren}}]{zhao_space-based_2023}%
  \BibitemOpen
  \bibfield  {author} {\bibinfo {author} {\bibfnamefont {T.}~\bibnamefont {Zhao}}, \bibinfo {author} {\bibfnamefont {R.}~\bibnamefont {Lyu}}, \bibinfo {author} {\bibfnamefont {H.}~\bibnamefont {Wang}}, \bibinfo {author} {\bibfnamefont {Z.}~\bibnamefont {Cao}},\ \bibnamefont {and}\ \bibinfo {author} {\bibfnamefont {Z.}~\bibnamefont {Ren}},\ }\href {\doibase 10.1038/s42005-023-01334-6} {\bibfield  {journal} {\bibinfo  {journal} {Commun. Phys.}\ }\textbf {\bibinfo {volume} {6}},\ \bibinfo {pages} {212} (\bibinfo {year} {2023}{\natexlab{b}})}\BibitemShut {NoStop}%
\bibitem [{\citenamefont {Wei}\ and\ \citenamefont {Huerta}(2020)}]{wei_gravitational_2020}%
  \BibitemOpen
  \bibfield  {author} {\bibinfo {author} {\bibfnamefont {W.}~\bibnamefont {Wei}}\ \bibnamefont {and}\ \bibinfo {author} {\bibfnamefont {E.~A.}\ \bibnamefont {Huerta}},\ }\href {\doibase 10.1016/j.physletb.2019.135081} {\bibfield  {journal} {\bibinfo  {journal} {Phys. Lett. B}\ }\textbf {\bibinfo {volume} {800}},\ \bibinfo {pages} {135081} (\bibinfo {year} {2020})}\BibitemShut {NoStop}%
\bibitem [{\citenamefont {Wei}(2021)}]{wei_applications_2021}%
  \BibitemOpen
  \bibfield  {author} {\bibinfo {author} {\bibfnamefont {W.}~\bibnamefont {Wei}},\ }\emph {\bibinfo {title} {Applications of Deep Learning for Gravitational Wave Physics}},\ \href {https://hdl.handle.net/2142/110444} {Ph.D. thesis},\ \bibinfo  {school} {University of Illinois Urbana-Champaign}, \bibinfo {address} {Champaign, IL, US} (\bibinfo {year} {2021})\BibitemShut {NoStop}%
\bibitem [{\citenamefont {Wang}\ \emph {et~al.}(2024{\natexlab{b}})\citenamefont {Wang}, \citenamefont {Zhou}, \citenamefont {Cao}, \citenamefont {Guo},\ and\ \citenamefont {Ren}}]{wang_waveformer_2024}%
  \BibitemOpen
  \bibfield  {author} {\bibinfo {author} {\bibfnamefont {H.}~\bibnamefont {Wang}}, \bibinfo {author} {\bibfnamefont {Y.}~\bibnamefont {Zhou}}, \bibinfo {author} {\bibfnamefont {Z.}~\bibnamefont {Cao}}, \bibinfo {author} {\bibfnamefont {Z.}~\bibnamefont {Guo}},\ \bibnamefont {and}\ \bibinfo {author} {\bibfnamefont {Z.}~\bibnamefont {Ren}},\ }\href {\doibase 10.1088/2632-2153/ad2f54} {\bibfield  {journal} {\bibinfo  {journal} {Mach. Learn.: Sci. Technol.}\ }\textbf {\bibinfo {volume} {5}},\ \bibinfo {pages} {015046} (\bibinfo {year} {2024}{\natexlab{b}})}\BibitemShut {NoStop}%
\bibitem [{\citenamefont {Dax}\ \emph {et~al.}(2021)\citenamefont {Dax}, \citenamefont {Green}, \citenamefont {Gair}, \citenamefont {Macke}, \citenamefont {Buonanno},\ and\ \citenamefont {Sch{\"o}lkopf}}]{dax_real-time_2021}%
  \BibitemOpen
  \bibfield  {author} {\bibinfo {author} {\bibfnamefont {M.}~\bibnamefont {Dax}}, \bibinfo {author} {\bibfnamefont {S.~R.}\ \bibnamefont {Green}}, \bibinfo {author} {\bibfnamefont {J.}~\bibnamefont {Gair}}, \bibinfo {author} {\bibfnamefont {J.~H.}\ \bibnamefont {Macke}}, \bibinfo {author} {\bibfnamefont {A.}~\bibnamefont {Buonanno}}, \bibnamefont {et~al.},\ }\href {\doibase 10.1103/PhysRevLett.127.241103} {\bibfield  {journal} {\bibinfo  {journal} {Phys. Rev. Lett.}\ }\textbf {\bibinfo {volume} {127}},\ \bibinfo {pages} {241103} (\bibinfo {year} {2021})}\BibitemShut {NoStop}%
\bibitem [{\citenamefont {Dax}\ \emph {et~al.}(2023)\citenamefont {Dax}, \citenamefont {Green}, \citenamefont {Gair}, \citenamefont {P{\"u}rrer}, \citenamefont {Wildberger}, \citenamefont {Macke}, \citenamefont {Buonanno},\ and\ \citenamefont {Sch{\"o}lkopf}}]{dax_neural_2023}%
  \BibitemOpen
  \bibfield  {author} {\bibinfo {author} {\bibfnamefont {M.}~\bibnamefont {Dax}}, \bibinfo {author} {\bibfnamefont {S.~R.}\ \bibnamefont {Green}}, \bibinfo {author} {\bibfnamefont {J.}~\bibnamefont {Gair}}, \bibinfo {author} {\bibfnamefont {M.}~\bibnamefont {P{\"u}rrer}}, \bibinfo {author} {\bibfnamefont {J.}~\bibnamefont {Wildberger}}, \bibnamefont {et~al.},\ }\href {\doibase 10.1103/PhysRevLett.130.171403} {\bibfield  {journal} {\bibinfo  {journal} {Phys. Rev. Lett.}\ }\textbf {\bibinfo {volume} {130}},\ \bibinfo {pages} {171403} (\bibinfo {year} {2023})}\BibitemShut {NoStop}%
\bibitem [{\citenamefont {Wildberger}\ \emph {et~al.}(2023)\citenamefont {Wildberger}, \citenamefont {Dax}, \citenamefont {Green}, \citenamefont {Gair}, \citenamefont {P{\"u}rrer}, \citenamefont {Macke}, \citenamefont {Buonanno},\ and\ \citenamefont {Sch{\"o}lkopf}}]{wildberger_adapting_2023}%
  \BibitemOpen
  \bibfield  {author} {\bibinfo {author} {\bibfnamefont {J.}~\bibnamefont {Wildberger}}, \bibinfo {author} {\bibfnamefont {M.}~\bibnamefont {Dax}}, \bibinfo {author} {\bibfnamefont {S.~R.}\ \bibnamefont {Green}}, \bibinfo {author} {\bibfnamefont {J.}~\bibnamefont {Gair}}, \bibinfo {author} {\bibfnamefont {M.}~\bibnamefont {P{\"u}rrer}}, \bibnamefont {et~al.},\ }\href {\doibase 10.1103/PhysRevD.107.084046} {\bibfield  {journal} {\bibinfo  {journal} {Phys. Rev. D}\ }\textbf {\bibinfo {volume} {107}},\ \bibinfo {pages} {084046} (\bibinfo {year} {2023})}\BibitemShut {NoStop}%
\bibitem [{\citenamefont {Langendorff}\ \emph {et~al.}(2023)\citenamefont {Langendorff}, \citenamefont {Kolmus}, \citenamefont {Janquart},\ and\ \citenamefont {Van Den~Broeck}}]{langendorff_normalizing_2023}%
  \BibitemOpen
  \bibfield  {author} {\bibinfo {author} {\bibfnamefont {J.}~\bibnamefont {Langendorff}}, \bibinfo {author} {\bibfnamefont {A.}~\bibnamefont {Kolmus}}, \bibinfo {author} {\bibfnamefont {J.}~\bibnamefont {Janquart}},\ \bibnamefont {and}\ \bibinfo {author} {\bibfnamefont {C.}~\bibnamefont {Van Den~Broeck}},\ }\href {\doibase 10.1103/PhysRevLett.130.171402} {\bibfield  {journal} {\bibinfo  {journal} {Phys. Rev. Lett.}\ }\textbf {\bibinfo {volume} {130}},\ \bibinfo {pages} {171402} (\bibinfo {year} {2023})}\BibitemShut {NoStop}%
\bibitem [{\citenamefont {Alvey}\ \emph {et~al.}(2023)\citenamefont {Alvey}, \citenamefont {Bhardwaj}, \citenamefont {Nissanke},\ and\ \citenamefont {Weniger}}]{alvey_what_2023}%
  \BibitemOpen
  \bibfield  {author} {\bibinfo {author} {\bibfnamefont {J.}~\bibnamefont {Alvey}}, \bibinfo {author} {\bibfnamefont {U.}~\bibnamefont {Bhardwaj}}, \bibinfo {author} {\bibfnamefont {S.}~\bibnamefont {Nissanke}},\ \bibnamefont {and}\ \bibinfo {author} {\bibfnamefont {C.}~\bibnamefont {Weniger}},\ }\href@noop {} {\enquote {\bibinfo {title} {What to do when things get crowded? {{Scalable}} joint analysis of overlapping gravitational wave signals},}\ } (\bibinfo {year} {2023}),\ \Eprint {http://arxiv.org/abs/2308.06318}{arXiv:2308.06318}\BibitemShut {NoStop}%
\bibitem [{\citenamefont {Ma}\ \emph {et~al.}(2024)\citenamefont {Ma}, \citenamefont {Zhou},\ and\ \citenamefont {Cao}}]{ma_gravitational_2024}%
  \BibitemOpen
  \bibfield  {author} {\bibinfo {author} {\bibfnamefont {C.}~\bibnamefont {Ma}}, \bibinfo {author} {\bibfnamefont {W.}~\bibnamefont {Zhou}},\ \bibnamefont {and}\ \bibinfo {author} {\bibfnamefont {Z.}~\bibnamefont {Cao}},\ }\href@noop {} {\enquote {\bibinfo {title} {Gravitational {{Wave Mixture Separation}} for {{Future Gravitational Wave Observatories Utilizing Deep Learning}}},}\ } (\bibinfo {year} {2024}),\ \Eprint {http://arxiv.org/abs/2407.13239}{arXiv:2407.13239}\BibitemShut {NoStop}%
\bibitem [{\citenamefont {Dal~Canton}\ \emph {et~al.}(2021)\citenamefont {Dal~Canton}, \citenamefont {Nitz}, \citenamefont {Gadre}, \citenamefont {Cabourn~Davies}, \citenamefont {{Villa-Ortega}}, \citenamefont {Dent}, \citenamefont {Harry},\ and\ \citenamefont {Xiao}}]{dal_canton_real-time_2021}%
  \BibitemOpen
  \bibfield  {author} {\bibinfo {author} {\bibfnamefont {T.}~\bibnamefont {Dal~Canton}}, \bibinfo {author} {\bibfnamefont {A.~H.}\ \bibnamefont {Nitz}}, \bibinfo {author} {\bibfnamefont {B.}~\bibnamefont {Gadre}}, \bibinfo {author} {\bibfnamefont {G.~S.}\ \bibnamefont {Cabourn~Davies}}, \bibinfo {author} {\bibfnamefont {V.}~\bibnamefont {{Villa-Ortega}}}, \bibnamefont {et~al.},\ }\href {\doibase 10.3847/1538-4357/ac2f9a} {\bibfield  {journal} {\bibinfo  {journal} {Astrophys. J.}\ }\textbf {\bibinfo {volume} {923}},\ \bibinfo {pages} {254} (\bibinfo {year} {2021})}\BibitemShut {NoStop}%
\bibitem [{\citenamefont {Boh{\'e}}\ \emph {et~al.}(2017)\citenamefont {Boh{\'e}}, \citenamefont {Shao}, \citenamefont {Taracchini}, \citenamefont {Buonanno}, \citenamefont {Babak}, \citenamefont {Harry}, \citenamefont {Hinder}, \citenamefont {Ossokine}, \citenamefont {P{\"u}rrer}, \citenamefont {Raymond}, \citenamefont {Chu}, \citenamefont {Fong}, \citenamefont {Kumar}, \citenamefont {Pfeiffer}, \citenamefont {Boyle}, \citenamefont {Hemberger}, \citenamefont {Kidder}, \citenamefont {Lovelace}, \citenamefont {Scheel},\ and\ \citenamefont {Szil{\'a}gyi}}]{bohe_improved_2017}%
  \BibitemOpen
  \bibfield  {author} {\bibinfo {author} {\bibfnamefont {A.}~\bibnamefont {Boh{\'e}}}, \bibinfo {author} {\bibfnamefont {L.}~\bibnamefont {Shao}}, \bibinfo {author} {\bibfnamefont {A.}~\bibnamefont {Taracchini}}, \bibinfo {author} {\bibfnamefont {A.}~\bibnamefont {Buonanno}}, \bibinfo {author} {\bibfnamefont {S.}~\bibnamefont {Babak}}, \bibnamefont {et~al.},\ }\href {\doibase 10.1103/PhysRevD.95.044028} {\bibfield  {journal} {\bibinfo  {journal} {Phys. Rev. D}\ }\textbf {\bibinfo {volume} {95}},\ \bibinfo {pages} {044028} (\bibinfo {year} {2017})}\BibitemShut {NoStop}%
\bibitem [{\citenamefont {Estell{\'e}s}\ \emph {et~al.}(2022)\citenamefont {Estell{\'e}s}, \citenamefont {Colleoni}, \citenamefont {{Garc{\'i}a-Quir{\'o}s}}, \citenamefont {Husa}, \citenamefont {Keitel}, \citenamefont {{Mateu-Lucena}}, \citenamefont {Planas},\ and\ \citenamefont {{Ramos-Buades}}}]{estelles_new_2022}%
  \BibitemOpen
  \bibfield  {author} {\bibinfo {author} {\bibfnamefont {H.}~\bibnamefont {Estell{\'e}s}}, \bibinfo {author} {\bibfnamefont {M.}~\bibnamefont {Colleoni}}, \bibinfo {author} {\bibfnamefont {C.}~\bibnamefont {{Garc{\'i}a-Quir{\'o}s}}}, \bibinfo {author} {\bibfnamefont {S.}~\bibnamefont {Husa}}, \bibinfo {author} {\bibfnamefont {D.}~\bibnamefont {Keitel}}, \bibnamefont {et~al.},\ }\href {\doibase 10.1103/PhysRevD.105.084040} {\bibfield  {journal} {\bibinfo  {journal} {Phys. Rev. D}\ }\textbf {\bibinfo {volume} {105}},\ \bibinfo {pages} {084040} (\bibinfo {year} {2022})}\BibitemShut {NoStop}%
\bibitem [{\citenamefont {Messina}\ \emph {et~al.}(2019)\citenamefont {Messina}, \citenamefont {Dudi}, \citenamefont {Nagar},\ and\ \citenamefont {Bernuzzi}}]{messina_quasi-55pn_2019}%
  \BibitemOpen
  \bibfield  {author} {\bibinfo {author} {\bibfnamefont {F.}~\bibnamefont {Messina}}, \bibinfo {author} {\bibfnamefont {R.}~\bibnamefont {Dudi}}, \bibinfo {author} {\bibfnamefont {A.}~\bibnamefont {Nagar}},\ \bibnamefont {and}\ \bibinfo {author} {\bibfnamefont {S.}~\bibnamefont {Bernuzzi}},\ }\href {\doibase 10.1103/PhysRevD.99.124051} {\bibfield  {journal} {\bibinfo  {journal} {Phys. Rev. D}\ }\textbf {\bibinfo {volume} {99}},\ \bibinfo {pages} {124051} (\bibinfo {year} {2019})}\BibitemShut {NoStop}%
\bibitem [{\citenamefont {Whelan}(2012)}]{whelan_visualization_2012}%
  \BibitemOpen
  \bibfield  {author} {\bibinfo {author} {\bibfnamefont {J.}~\bibnamefont {Whelan}},\ }\href {https://dcc.ligo.org/LIGO-T1100431/public} {\emph {\bibinfo {title} {Visualization of {{Antenna Pattern Factors}} via {{Projected Detector Tensors}}}}},\ \bibinfo {type} {Tech. Rep.}\ \bibinfo {number} {LIGO-T1100431-v2}\ (\bibinfo {year} {2012})\BibitemShut {NoStop}%
\bibitem [{\citenamefont {Chen}\ \emph {et~al.}(2020)\citenamefont {Chen}, \citenamefont {Mao},\ and\ \citenamefont {Liu}}]{chen_dual-path_2020}%
  \BibitemOpen
  \bibfield  {author} {\bibinfo {author} {\bibfnamefont {J.}~\bibnamefont {Chen}}, \bibinfo {author} {\bibfnamefont {Q.}~\bibnamefont {Mao}},\ \bibnamefont {and}\ \bibinfo {author} {\bibfnamefont {D.}~\bibnamefont {Liu}},\ }in\ \href {\doibase 10.21437/Interspeech.2020-2205} {\emph {\bibinfo {booktitle} {Interspeech 2020}}}\ (\bibinfo  {publisher} {ISCA},\ \bibinfo {year} {2020})\ pp.\ \bibinfo {pages} {2642--2646}\BibitemShut {NoStop}%
\bibitem [{\citenamefont {Luo}\ \emph {et~al.}(2020)\citenamefont {Luo}, \citenamefont {Chen},\ and\ \citenamefont {Yoshioka}}]{luo_dual-path_2020}%
  \BibitemOpen
  \bibfield  {author} {\bibinfo {author} {\bibfnamefont {Y.}~\bibnamefont {Luo}}, \bibinfo {author} {\bibfnamefont {Z.}~\bibnamefont {Chen}},\ \bibnamefont {and}\ \bibinfo {author} {\bibfnamefont {T.}~\bibnamefont {Yoshioka}},\ }in\ \href {\doibase 10.1109/ICASSP40776.2020.9054266} {\emph {\bibinfo {booktitle} {{{ICASSP}} 2020 - 2020 {{IEEE International Conference}} on {{Acoustics}}, {{Speech}} and {{Signal Processing}} ({{ICASSP}})}}}\ (\bibinfo {year} {2020})\ pp.\ \bibinfo {pages} {46--50}\BibitemShut {NoStop}%
\bibitem [{\citenamefont {Zhu}(2020)}]{junzhe_multi-decoder_2020}%
  \BibitemOpen
  \bibfield  {author} {\bibinfo {author} {\bibfnamefont {J.}~\bibnamefont {Zhu}},\ }\emph {\bibinfo {title} {Multi-Decoder {{DPRNN}} High Accuracy Source Counting and Separation}},\ \href {https://hdl.handle.net/2142/109201} {\bibinfo {type} {Senior}},\ \bibinfo  {school} {University of Illinois at Urbana-Champaign}, \bibinfo {address} {Champaign, IL, US} (\bibinfo {year} {2020}),\ \Eprint {http://arxiv.org/abs/2011.12022}{arXiv:2011.12022}\BibitemShut {NoStop}%
\bibitem [{\citenamefont {Dong}\ \emph {et~al.}(2024)\citenamefont {Dong}, \citenamefont {Li}, \citenamefont {Tao}, \citenamefont {Jiang}, \citenamefont {Zhang}, \citenamefont {Li}, \citenamefont {Su}, \citenamefont {Zhang},\ and\ \citenamefont {Xu}}]{dong_fan_2024}%
  \BibitemOpen
  \bibfield  {author} {\bibinfo {author} {\bibfnamefont {Y.}~\bibnamefont {Dong}}, \bibinfo {author} {\bibfnamefont {G.}~\bibnamefont {Li}}, \bibinfo {author} {\bibfnamefont {Y.}~\bibnamefont {Tao}}, \bibinfo {author} {\bibfnamefont {X.}~\bibnamefont {Jiang}}, \bibinfo {author} {\bibfnamefont {K.}~\bibnamefont {Zhang}}, \bibnamefont {et~al.},\ }\href@noop {} {\enquote {\bibinfo {title} {{{FAN}}: {{Fourier Analysis Networks}}},}\ } (\bibinfo {year} {2024}),\ \Eprint {http://arxiv.org/abs/2410.02675}{arXiv:2410.02675}\BibitemShut {NoStop}%
\bibitem [{\citenamefont {Vincent}\ \emph {et~al.}(2006)\citenamefont {Vincent}, \citenamefont {Gribonval},\ and\ \citenamefont {Fevotte}}]{vincent_performance_2006}%
  \BibitemOpen
  \bibfield  {author} {\bibinfo {author} {\bibfnamefont {E.}~\bibnamefont {Vincent}}, \bibinfo {author} {\bibfnamefont {R.}~\bibnamefont {Gribonval}},\ \bibnamefont {and}\ \bibinfo {author} {\bibfnamefont {C.}~\bibnamefont {Fevotte}},\ }\href {\doibase 10.1109/TSA.2005.858005} {\bibfield  {journal} {\bibinfo  {journal} {IEEE Trans. Audio Speech Lang. Process.}\ }\textbf {\bibinfo {volume} {14}},\ \bibinfo {pages} {1462} (\bibinfo {year} {2006})}\BibitemShut {NoStop}%
\bibitem [{\citenamefont {Pariente}\ \emph {et~al.}(2020)\citenamefont {Pariente}, \citenamefont {Cornell}, \citenamefont {Cosentino}, \citenamefont {Sivasankaran}, \citenamefont {Tzinis}, \citenamefont {Heitkaemper}, \citenamefont {Olvera}, \citenamefont {St{\"o}ter}, \citenamefont {Hu}, \citenamefont {{Mart{\'i}n-Do{\~n}as}}, \citenamefont {Ditter}, \citenamefont {Frank}, \citenamefont {Deleforge},\ and\ \citenamefont {Vincent}}]{pariente_asteroid_2020}%
  \BibitemOpen
  \bibfield  {author} {\bibinfo {author} {\bibfnamefont {M.}~\bibnamefont {Pariente}}, \bibinfo {author} {\bibfnamefont {S.}~\bibnamefont {Cornell}}, \bibinfo {author} {\bibfnamefont {J.}~\bibnamefont {Cosentino}}, \bibinfo {author} {\bibfnamefont {S.}~\bibnamefont {Sivasankaran}}, \bibinfo {author} {\bibfnamefont {E.}~\bibnamefont {Tzinis}}, \bibnamefont {et~al.},\ }in\ \href {\doibase 10.21437/Interspeech.2020-1673} {\emph {\bibinfo {booktitle} {Interspeech 2020}}}\ (\bibinfo  {publisher} {ISCA},\ \bibinfo {year} {2020})\ pp.\ \bibinfo {pages} {2637--2641}\BibitemShut {NoStop}%
\bibitem [{\citenamefont {Wu}\ \emph {et~al.}(2022)\citenamefont {Wu}, \citenamefont {Hu}, \citenamefont {Liu}, \citenamefont {Zhou}, \citenamefont {Wang},\ and\ \citenamefont {Long}}]{wu_timesnet_2022}%
  \BibitemOpen
  \bibfield  {author} {\bibinfo {author} {\bibfnamefont {H.}~\bibnamefont {Wu}}, \bibinfo {author} {\bibfnamefont {T.}~\bibnamefont {Hu}}, \bibinfo {author} {\bibfnamefont {Y.}~\bibnamefont {Liu}}, \bibinfo {author} {\bibfnamefont {H.}~\bibnamefont {Zhou}}, \bibinfo {author} {\bibfnamefont {J.}~\bibnamefont {Wang}}, \bibnamefont {et~al.},\ }in\ \href {https://openreview.net/forum?id=ju_Uqw384Oq} {\emph {\bibinfo {booktitle} {The {{Eleventh International Conference}} on {{Learning Representations}}}}}\ (\bibinfo {year} {2022})\BibitemShut {NoStop}%
\bibitem [{\citenamefont {Wang}\ \emph {et~al.}(2024{\natexlab{c}})\citenamefont {Wang}, \citenamefont {Wu}, \citenamefont {Dong}, \citenamefont {Liu}, \citenamefont {Long},\ and\ \citenamefont {Wang}}]{wang_deep_2024}%
  \BibitemOpen
  \bibfield  {author} {\bibinfo {author} {\bibfnamefont {Y.}~\bibnamefont {Wang}}, \bibinfo {author} {\bibfnamefont {H.}~\bibnamefont {Wu}}, \bibinfo {author} {\bibfnamefont {J.}~\bibnamefont {Dong}}, \bibinfo {author} {\bibfnamefont {Y.}~\bibnamefont {Liu}}, \bibinfo {author} {\bibfnamefont {M.}~\bibnamefont {Long}}, \bibnamefont {et~al.},\ }\href@noop {} {\enquote {\bibinfo {title} {Deep {{Time Series Models}}: {{A Comprehensive Survey}} and {{Benchmark}}},}\ } (\bibinfo {year} {2024}{\natexlab{c}}),\ \Eprint {http://arxiv.org/abs/2407.13278}{arXiv:2407.13278}\BibitemShut {NoStop}%
\bibitem [{\citenamefont {Kingma}\ and\ \citenamefont {Ba}(2015)}]{kingma_adam_2015}%
  \BibitemOpen
  \bibfield  {author} {\bibinfo {author} {\bibfnamefont {D.~P.}\ \bibnamefont {Kingma}}\ \bibnamefont {and}\ \bibinfo {author} {\bibfnamefont {J.}~\bibnamefont {Ba}},\ }in\ \href@noop {} {\emph {\bibinfo {booktitle} {3rd International Conference on Learning Representations, {{ICLR}} 2015, San Diego, {{CA}}, {{USA}}}}},\ \bibinfo {editor} {edited by\ \bibinfo {editor} {\bibfnamefont {Y.}~\bibnamefont {Bengio}}\ \bibnamefont {and}\ \bibinfo {editor} {\bibfnamefont {Y.}~\bibnamefont {LeCun}}}\ (\bibinfo {year} {2015})\ \Eprint {http://arxiv.org/abs/1412.6980}{arXiv:1412.6980}\BibitemShut {NoStop}%
\bibitem [{\citenamefont {Liu}\ \emph {et~al.}(2024)\citenamefont {Liu}, \citenamefont {Cao},\ and\ \citenamefont {Zhu}}]{liu_effectiv_2024}%
  \BibitemOpen
  \bibfield  {author} {\bibinfo {author} {\bibfnamefont {X.}~\bibnamefont {Liu}}, \bibinfo {author} {\bibfnamefont {Z.}~\bibnamefont {Cao}},\ \bibnamefont {and}\ \bibinfo {author} {\bibfnamefont {Z.-H.}\ \bibnamefont {Zhu}},\ }\href {\doibase 10.1088/1361-6382/ad72ca} {\bibfield  {journal} {\bibinfo  {journal} {Class. Quantum Gravity}\ }\textbf {\bibinfo {volume} {41}},\ \bibinfo {pages} {195019} (\bibinfo {year} {2024})}\BibitemShut {NoStop}%
\bibitem [{\citenamefont {Cao}\ and\ \citenamefont {Han}(2017)}]{cao_waveform_2017}%
  \BibitemOpen
  \bibfield  {author} {\bibinfo {author} {\bibfnamefont {Z.}~\bibnamefont {Cao}}\ \bibnamefont {and}\ \bibinfo {author} {\bibfnamefont {W.-B.}\ \bibnamefont {Han}},\ }\href {\doibase 10.1103/PhysRevD.96.044028} {\bibfield  {journal} {\bibinfo  {journal} {Phys. Rev. D}\ }\textbf {\bibinfo {volume} {96}},\ \bibinfo {pages} {044028} (\bibinfo {year} {2017})}\BibitemShut {NoStop}%
\end{thebibliography}%

\end{document}